\documentclass[11pt]{article}

\input{preface/includes}

\usepackage{amsmath}
\usepackage{amsthm}
\usepackage{amssymb}
\usepackage{authblk}
\usepackage{bm}
\usepackage{mathtools}

\newlength{\arrow}
\newcommand*{\goesto}[1]{\xrightarrow{\mathmakebox[\arrow]{#1}}}
\DeclareMathOperator*{\setsup}{sup}

\DeclarePairedDelimiter\ceil{\lceil}{\rceil}
\newcommand{\td}{[0,\infty)}

\newcommand{\bdelta}{{\bm{\delta}}}
\newcommand{\xn}{\mathbf{x}_0}
\newcommand{\bu}{\mathbf{u}}
\newcommand{\buhat}{\hat{\mathbf{u}}}
\newcommand{\hhat}{\hat{h}}
\newcommand{\khat}{\hat{k}}
\newcommand{\bv}{\mathbf{v}}
\newcommand{\bx}{\mathbf{x}}
\newcommand{\bc}{\mathbf{c}}
\newcommand{\bchat}{\hat{\mathbf{c}}}
\newcommand{\br}{\mathbf{r}}
\newcommand{\bp}{\mathbf{p}}
\newcommand{\xnhat}{\hat{\mathbf{x}}_0}
\newcommand{\bxhat}{\hat{\mathbf{x}}}
\newcommand{\yq}{y_q}

\newcommand{\Yq}{Y_q}
\newcommand{\Zq}{Z_q}
\newcommand{\zq}{z_q}
\newcommand{\Xc}{X_c}

\newcommand{\xcs}{x_c^\ast}
\newcommand{\xc}{x_c^\ast}
\newcommand{\Xa}{X_a}
\newcommand{\xas}{x_a^\ast}
\newcommand{\xa}{x_a^\ast}
\newcommand{\xrs}{x_r^\ast}
\newcommand{\xr}{x_r^\ast}
\newcommand{\Xr}{X_r}
\newcommand{\pyq}{p(Y_q)}
\newcommand{\pzq}{p(Z_q)}
\newcommand{\yqbar}{\overline{y}_q}
\newcommand{\Yqbar}{\overline{Y}_q}
\newcommand{\zqbar}{\overline{z}_q}

\newcommand{\Zqbar}{\overline{Z}_q}

\newcommand{\Nhat}{\hat{N}}

\newcommand{\false}{\textbf{\text{false}}}
\newcommand{\true}{\textbf{\text{true}}}
\newcommand{\D}[1]{\frac{d#1}{dt}}
\newcommand{\norm}[1]{\lVert #1 \rVert}

\newcommand{\deltaof}[1]{\widehat{\Delta}(I,#1)}

\newcommand*{\bS}{\mathbf{S}}
\newcommand*{\supX}{{{(X)}}}
\newcommand*{\Xbar}{{\overline{X}}}
\newcommand*{\xbar}{\overline{x}}
\newcommand*{\Ybar}{{\overline{Y}}}
\newcommand*{\Zbar}{{\overline{Z}}}
\newcommand*{\Xhat}{{\widehat{X}}}
\newcommand*{\ys}{y_s}

\numberwithin{equation}{section}
\newtheorem{theorem}{Theorem}[section]
\newtheorem{lemma}[theorem]{Lemma}
\newtheorem{corollary}[theorem]{Corollary}
\newtheorem{construction}[theorem]{Construction}

\newtheorem{observation}[theorem]{Observation}
\newtheorem*{observation*}{Observation}

\usepackage[capitalise]{cleveref}
\crefname{sec}{Sect.}{Sect.}
\Crefname{sec}{Section}{Sections}
\crefname{listing}{\lstlistingname}{\lstlistingname}
\Crefname{listing}{Listing}{Listings}

\begin{document}

\title{
    Robust Biomolecular Finite Automata
    \thanks{This research was supported in part by National Science Foundation Grants 1247051 and 1545028.}
}


\author[1]{Titus H. Klinge}
\author[2]{James I. Lathrop}
\author[2]{Jack H. Lutz}
\affil[1]{Carleton College, Northfield, MN 55057 USA\\
    \texttt{tklinge@carleton.edu}}
\affil[2]{Iowa State University, Ames, IA 50011 USA\\
    \texttt{\{jil,lutz\}@iastate.edu}}

\date{}


\maketitle

    \begin{abstract}
	We present a uniform method for translating an arbitrary nondeterministic finite automaton (NFA) into a deterministic mass action \emph{input/output chemical reaction network} (\emph{I/O CRN}) that simulates it.
    The I/O CRN receives its input as a continuous time signal consisting of concentrations of chemical species that vary to represent the NFA's input string in a natural way.
    The I/O CRN exploits the inherent parallelism of chemical kinetics to simulate the NFA in \emph{real time} with a number of chemical species that is \emph{linear} in the size of the NFA.
    We prove that the simulation is correct and that it is robust with respect to perturbations of the input signal, the initial concentrations of species, the output (decision), and the rate constants of the reactions of the I/O CRN.

    \vspace*{1em}
\end{abstract}

    \section{Introduction}\label{sec:intro}
Molecular programming combines computer science principles with the information processing capabilities of DNA and other biomolecules in order to control the structure and behavior of matter at the nanoscale.
Molecular programming has its origin in Seeman's development of DNA nanotechnology in the 1980s~\cite{jSeem82} (indeed, \qq{molecular programming} and \qq{DNA nanotechnology} are still nearly synonymous), but the field has made progress in the present century at a rate whose increase is reminiscent of Moore's law.
The achievements of molecular programming are far too numerous to survey here, but they include the self-assembly of virtually any two- or three-dimensional nanoscale structure that one wants to prescribe~\cite{jRoth06,jDMTVCS09,jHPNDLY11,jKOSY12,jWeDaYi12}, DNA strand displacement networks that simulate logic circuits and neural networks~\cite{jQiaWin11,jQiaWin11a,jQiWiBr11}, and molecular robots that perform various functions while either walking on nanoscale tracks or floating free in solution~\cite{jSmit10,jWYEWT11,jZhaSee11,jCSHELM12,jDoBaCh12,cDKTT13,jYTMSN00}.
All this has been achieved in real laboratory experiments, and applications to synthetic biology, medicine, and computer electronics are envisioned.
Theoretical progress includes demonstrations that various molecular programming paradigms are, in principle, Turing universal~\cite{oWinf98,jAAER07,jSCWB08,oCSWB09,cDLPSSW12,jWood15,cFLBP17},
thereby indicating that the full generality and creativity of algorithmic computation may be deployed in molecular and biological arenas.

Our objective in this paper is to begin mitigating the \qq{in principle} of the preceding sentence.
This is important for two reasons.
First, although such theoretical results are steps in the right direction, processes that require unrealistically precise control of unrealistically large numbers of molecules simply cannot be implemented.
Second, processes that can be implemented, but only with inordinately precise control of parameters are inherently unreliable and hence inherently unsafe in many envisioned applications.
Our objective here is thus to identify a class of computations that can be implemented \emph{robustly} in the molecular world, \ie, implemented in such a way that they will \emph{provably} perform correctly, even when crucial parameters are perturbed by small amounts.
Future research can then strive to enhance this robustness and to extend the class of computations that enjoy it.

In this paper we give a uniform method for translating nondeterministic finite automata to chemical reaction networks that implement them robustly.
Nondeterministic finite automata (NFAs) are over half a century old~\cite{jRabSco59} and far from Turing universal, but they have many applications and remain an active research topic~\cite{cBonPou13,jHenRas15,jBonPou15}.
Applications of NFAs that are likely to extend to molecular programming include their uses in monitoring and parsing large data streams and in implementing and verifying secure and/or safe communication protocols.
Chemical reaction networks (CRNs) are also over half a century old~\cite{jAris65}.
Their role in molecular programming did not become fully apparent until recently, when Soloveichik, Seelig, and Winfree~\cite{jSoSeWi10} showed that there is a systematic method for translating an arbitrary CRN, which is an abstract mathematical object, into a set of DNA strands and complexes that simulates the CRN via toehold-mediated strand displacement.
This method has been refined and formulated as a compiler~\cite{jCDSPCS13,cBSJDTW17},
and CRNs are now the programming language of choice for many molecular programming investigations.
The two most widely used semantics (operational meanings) for CRNs are deterministic mass action semantics and stochastic mass action semantics.
In this paper we use deterministic mass action, which implies that the state of a CRN at any time is determined by the \emph{real-valued concentrations} of its molecular species at that time.

An NFA is a real-time device that reads its input string \emph{sequentially}, left to right, changing states appropriately in response to each symbol prior to reading the next symbol.
Accordingly, we translate each NFA to an \emph{input/output} \emph{CRN} (\emph{I/O CRN}), which is a CRN that receives the NFA's input string formatted as a continuous time \emph{concentration signal} consisting of concentrations of input species that vary to represent the input string in a natural way.
(Concentration signals are likely to be useful in other molecular programming contexts, e.g., in modularizing CRN constructions.) Using the inherent parallelism of chemical kinetics, our I/O CRN implements the NFA in \emph{real time}, processing each input symbol before the next one arrives, and it does so with a number of molecular species that is \emph{linear} in the size of the NFA that it implements.
Specifically, if the NFA has \( q \) states, \( s \) symbols, and \( d \) transitions, then our I/O CRN consists of two modular components.
The first module is a preprocessor that transforms the input concentration signals into approximate square waves and consists of \( (s+2)(n+4) \) species and \( 2(s+2)(n+2) \) reactions, where \( n \) is logarithmic in \( q \).
The second module, which actually simulates the NFA, has \( 4q+s+2 \) species and \( 5q+d \) reactions.
The compiler of~\cite{jCDSPCS13} would then translate these modules into DNA gates and strands for a strand displacement network consisting of \( 4(s+2)(n+2) \) gates and \( 7(s+2)(n+1) \)
strands for the first module, and \( 10q + 2d \) gates and \( 24q + 5d \) strands for the second module.

Our translation thus appears to make small NFAs implementable in laboratories now and NFAs of modest size implementable in the near future.

Most importantly, our I/O CRN's correct implementation of the NFA is robust with respect to small perturbations of four things, namely, its input signal, the initial concentrations of its species, its output signal (acceptance or rejection of its input string), and the rate constants of its reactions.
One key to achieving this robustness is a signal restoration technique akin to the approximate majority algorithm of~\cite{jAnAsEi08a,jCarCsi12,jCard14}.

It should be noted that our NFA construction also supports infinite sequences of symbols, and so our results can be immediately applied to omega automata.

The rest of this paper is organized as follows.
\Cref{sec:io_crns} defines the I/O CRN model, an extension of the CRN model.
\Cref{sec:req_and_robust} introduces a specific notion of a requirement and then uses such requirements to specify robustness properties of I/O CRNs.
\Cref{sec:preproc} gives a construction and theorem for a CRN module that enhances input signals by removing noise.
\Cref{sec:nfa_sim} presents the main result of the paper, a construction for the robust simulation of an NFA using I/O CRNs and the proof that the construction is correct.
Some concluding remarks are given in \Cref{sec:conclusion}.  Finally, detailed technical proofs of certain lemmas are provided in the appendices.

    \section{Input/Output Reaction Networks}\label{sec:io_crns}
The chemical reaction network model used here must, like the sequential automata that it simulates, have a provision for reading its input over a period of time, processing early parts of the input before later parts become available.
This section describes a chemical reaction network model with such a provision.
Inputs are read as \emph{concentration signals}, which consist of concentrations of designated input species that vary over time under external control.
This model takes its name from the fact that its deterministic mass action semantics, developed below, is a special case of the ``input/output systems'' of control theory.

Formally, we fix a countably infinite set \( \mathbf{S} = \{X_0, X_1, \ldots \} \), whose elements we call \emph{species}.
Informally, we regard each species as an abstract name of a type of molecule, and we avoid excessive subscripts by writing elements of \( \mathbf{S} \) in other ways, \eg, \( X \), \( Y \), \( Z \), \( \widehat{X} \), \( \widetilde{X} \), \etc.

A \emph{reaction} over a finite set \( S \subseteq \mathbf{S} \) is formally a triple
\[
    \rho = (\br, \bp, k) \in \mathbb{N}^S \times \mathbb{N}^S \times (0, \infty),
\]
where \( \mathbb{N}^S \) is the set of functions from \( S \) into \( \mathbb{N} = \{0, 1, 2, \ldots \} \), and \( \br \neq \bp \).
Since \( S \) is finite, it is natural to also regard elements of \( \mathbb{N}^S \) as vectors.
Given such a reaction \( \rho \), we write \( \br(\rho) = \br \), \( \bp(\rho) = \bp \), and \( k(\rho) = k \), and we call these three things the \emph{reactant vector}, the \emph{product vector}, and the \emph{rate constant}, respectively, of the reaction \( \rho \).
The species in the \emph{support set} \( \text{supp}(\br) = \{ X \in S \mid \br(X) > 0 \} \) are the \emph{reactants} of \( \rho \), and the species in \( \text{supp}(\bp) \) are the \emph{products} of \( \rho \).

We usually write reactions in a more intuitive, chemistry-like notation.
For example, if \( S = \{X, Y, Z\} \), then we write
\[
    X + Y \goesto{k} X + 2Z
\]
for the reaction \( (\br, \bp, k) \), where \( \br \), \( \bp : S \rightarrow \mathbb{N} \) are defined by \( \br(X) = \br(Y) = 1 \), \( \br(Z) = 0 \), \( \bp(X) = 1 \), \( \bp(Y) = 0 \), and \( \bp(Z) = 2 \).

The \emph{net effect} of a reaction \( \rho \) is the (nonzero) vector \( \Delta\rho = \bp(\rho) - \br(\rho) \in \mathbb{Z}^S \).
A species \( X \) satisfying \( \br(\rho)(X) = \bp(\rho)(X) > 0 \), as in the example above, is called a \emph{catalyst} of the reaction \( \rho \).

An \emph{input/output chemical reaction network} (\emph{I/O CRN}) is an ordered triple \( N=(U,R,S) \), where
\( U,S\subseteq\mathbf{S} \) are finite;
\( U\cap S=\emptyset \);
\( R \) is a finite set of reactions over \( U\cup S \);
and species in \( U \) only appear as catalysts in \( R \).
Elements of \( S \) are called \emph{state species}, or \emph{operating species}, of \( N \).
Elements of \( U \) are called \emph{input species} of \( N \).

Given a finite set \( W\subseteq\textbf{S} \) of species, we define the \( W \)-\emph{signal space} to be the set \( C[W] = C^\omega(\td,\td^W) \), where \( C^\omega(\mathcal{X},\mathcal{Y}) \) is the set of real analytic functions from \( \mathcal{X} \) to \( \mathcal{Y} \).
A function \( \mathbf{w}\in C[W] \) is a \emph{concentration signal} that specifies the \emph{concentration} \( \mathbf{w}(t)(Y)\in\td \) of each species \( Y\in W \) at each time \( t\in\td \).

For sets \( W,W'\subseteq\mathbf{S} \), we also use the setc
\( C[W,W']=C(\td^W,\td^{W'}) \).

Intuitively, an I/O CRN \( N=(U,R,S) \) is a system that transforms an input signal \( \bu\in C[U] \) to an output signal \( \bv \).
We now make this intuition precise.

A \emph{context} of an I/O CRN \( N=(U,R,S) \) is an ordered triple \( \bc=(\bu,V,h) \), where \( \bu\in C[U] \) is an \emph{input signal}, \( V\subseteq S \) is a set of \emph{output species}, and \( h\in C[S\cup U,V] \) is an \emph{output function}.
We write \( \mathcal{C}_N \) for the set of contexts of \( N \).

The \emph{deterministic mass action semantics} (or \emph{deterministic mass action kinetics}) of an I/O CRN \( N \) specifies how \( N \) behaves in a context \( (\bu,V,h) \).

Let \( N=(U,R,S) \) be an I/O CRN.
A \emph{state} of \( N \) is a vector \( \bx\in\td^S \); an \emph{input state} of \( N \) is a vector \( \bu\in\td^U \); and a \emph{global state} of \( N \) is a vector \( (\bx,\bu)\in\td^{S\cup U} \), where \( \bx \) is a state of \( N \) and \( \bu \) is an input state of \( N \).
(Our double usage of the notation \( \bu \) for a single input state and also for a function specifying a time-varying input state \( \bu(t) \) is deliberate and minimizes obfuscation.
The same holds for \( \bx \) and \( \bx(t) \) below.)
For each reaction \( \rho\in R \) and each \( (\bx,\bu)\in\td^{S\cup U} \), the (\emph{deterministic mass action}) \emph{rate of} \( \rho \) \emph{in the global state} \( (\bx,\bu) \) is
\begin{equation}\label{eq:reaction_rate}
	\text{rate}_{\bx,\bu}(\rho) = k(\rho) (\bx,\bu)^{\br(\rho)},
\end{equation}
where \( (\bx,\bu)^{\br(\rho)} \) is the product, for all \( Y \in S \cup U \), of c
\( (\bx,\bu)(Y)^{\br(\rho)(Y)} \).
For example, if \( \rho \) is the reaction \( X+Y\goesto{k} X+2Z \), where \( X\in U \) and \( Y,Z\in S \), then \( \text{rate}_{\bx,\bu}(\rho)=k\bu(X)\bx(Y) \).
Intuitively, the frequency with which an \( X \) and a \( Y \) react with one another is proportional to \( \bu(X)\bx(Y) \), and the constant of proportionality \( k \) summarizes other factors, not depending on \( (\bx,\bu) \) (\eg temperature, salinity of solution, properties of \( X \) and \( Y \)), that also govern the rate at which \( \rho \) occurs.

For each state species \( Y\in S \), we define the \emph{deterministic mass action function} \( F_Y:\td^{S\cup U}\rightarrow\mathbb{R} \) by
\begin{equation}
	F_Y(\bx,\bu)=\sum_{\rho\in R}\text{rate}_{\bx,\bu}(\rho)\Delta\rho(Y)
\end{equation}
for all \( \bx\in\td^S \) and \( \bu\in\td^U \).
Then \( F_Y(\bx,\bu) \) is the total rate at which the concentration of \( Y \) is changing in the global state \( (\bx,\bu) \).
Now let \( (\bu,V,h) \) be a context of the I/O CRN \( N \).
Then \( \bu(t)(X) \) is the concentration of each input species \( X\in U \) at each time \( t\in\td \).
Hence, if the state of \( N \) is \( \bx(t)\in\td^S \) at time \( t \), then the concentration of each state species \( Y \) must obey the ordinary differential equation (ODE)
\begin{equation}\label{eq:mass_action_ode}
	y'(t)=F_Y(\bx(t),\bu(t)).
\end{equation}
If we let \( \mathcal{E}_Y \) be the ODE~\eqref{eq:mass_action_ode} for each \( Y\in S \), then the \emph{deterministic mass action system} of the I/O CRN \( N \) is the coupled system
\begin{equation}\label{eq:mass_action_system}
	(\mathcal{E}_Y\mid Y\in S)
\end{equation}
of ODEs.
If we define the vector-valued function \( F:\td^{S\cup U}\rightarrow \mathbb{R}^S \) by
\begin{equation}
	F(\bx,\bu)=(F_Y(\bx,\bu)\mid Y\in S)
\end{equation}
for all \( \bx\in\td^S \) and \( \bu\in\td^U \), then the mass action system~\eqref{eq:mass_action_system} can also be written in the vector form
\begin{equation}\label{eq:mass_action_system_vector}
	\bx'(t)=F(\bx(t),\bu(t)).
\end{equation}

The I/O CRN \( N \) is initialized to a state \( \xn\in\td^S \) at time 0 in the context \( (\bu,V,h) \), and this state then evolves according to the mass action system~\eqref{eq:mass_action_system_vector}.
The \emph{deterministic mass action initial value problem} (\emph{IVP}) of \( N \) in the context \( (\bu,V,h) \) with the initial state \( \xn \) is thus the initial value problem consisting of the mass action system~\eqref{eq:mass_action_system_vector} together with the initial value condition
\begin{equation}
	y(0) = \bx_0(Y) \text{  for each \( Y \in S \)}.
\end{equation}
By the standard existence-uniqueness theory for ODEs~\cite{oApos69,oTesc12}, this mass action IVP has a solution \( \bx(t) \) that is defined for all \( t \in [0, b) \) for some \( b \in (0, \infty] \), and this solution is unique.
It is not difficult to show, then, that \( \bx(t) \in [0, \infty)^S \) holds for all \( t \in [0, b) \), \ie, that concentrations remain nonnegative.
The I/O CRNs defined in this paper are all very well behaved, so that \( b = \infty \), \ie, \( \bx(t) \) is well defined for all \( t \in [0, \infty) \) and all input signals and initial values considered here.

In the context \( (\bu,V,h) \) of \( N \), the observed output of \( N \) is given by the output function \( h:\td^{S\cup U}\rightarrow\td^V \).
In most cases, this function \( h \) is some approximation, due to experimental error, of the zero-error projection function \( h_0:\td^{S\cup U}\rightarrow\td^V \) defined by
\begin{equation}\label{eq:zero_error_projection_function}
	h_0(\bx,\bu)(Y)=\bx(Y)
\end{equation}
for all \( Y\in V \).
If \( \bx(t) \) is defined as in the preceding paragraph, then the \emph{output signal} of the I/O CRN \( N \) in the context \( \bc=(\bu,h) \) with the initial state \( \xn \) is the (continuous) function \( N_{\bc,\xn}:\td\rightarrow\td^V \) defined by \( N_{\bc,\xn}(t)=h(\bx(t),\bu(t)) \)
for all \( t\in\td \).

In the language of control theory~\cite{oAstMur08,oDelMur14}, an \emph{input/output system} is a system of the form~\eqref{eq:mass_action_system_vector}, where \( \bx(t) \) and \( \bu(t) \) range over more general state spaces \( \mathcal{X} \) and \( \mathcal{U} \), together with a function \( h:\mathcal{X}\times\mathcal{U}\rightarrow\mathcal{V} \) for some space \( \mathcal{V} \) of values.
The input signal \( \bu \) is often called a \emph{control signal}, and the output function \( h \) is often called a \emph{measurement function}.

In most papers, a \emph{chemical reaction network} (\emph{CRN}) is an ordered pair \( N=(S,R) \) such that \( (\emptyset, R, S) \) is an I/O CRN as defined here.
Such CRNs are \emph{autonomous} in the two equivalent senses that
(i) the system~\eqref{eq:mass_action_system_vector} has the simpler form
\begin{equation}\label{eq:mass_action_system_simple}
	\bx'(t)=F(\bx(t)),
\end{equation}
the right-hand side of which only depends on the time \( t \) indirectly, via the state \( \bx(t) \); and
(ii) once the initial state \( \bx(0) \) is determined, the CRN's state evolves according to~\eqref{eq:mass_action_system_simple}, without further outside influence.
It is clear by inspection of~\eqref{eq:reaction_rate}-\eqref{eq:mass_action_system_vector} that the deterministic mass action system~\eqref{eq:mass_action_system_simple} of an autonomous CRN is \emph{polynomial}, meaning that the components of the vector \( F(\bx(t)) \) are polynomial in the components \( y(t) \) of \( \bx(t) \).
In contrast, the I/O CRNs considered in the present paper have mass action systems~\eqref{eq:mass_action_system_vector} that are neither autonomous nor polynomial.

Further discussions of chemical reaction networks with deterministic  mass action semantics appear in~\cite{oErdTot89,oEpsPoj98,oGuna03,oLent15}.

We conclude this section by noting that I/O CRNs offer a natural means for modularizing constructions.
It is often convenient to write the components of an I/O CRN \( N=(U,R,S) \) as \( U[N]=U \), \( R[N]=R \), and \( S[N]=S \).
The \emph{join} of a finite family \( \mathcal{N} \) of I/O CRNs is the I/O CRN
\[
    \bigsqcup\mathcal{N}=(U^\ast\setminus S^\ast, R^\ast, S^\ast),
\]
where \( U^\ast=\bigcup_{N\in\mathcal{N}}U[N] \), \( R^\ast=\bigcup_{N\in\mathcal{N}}R[N] \), and \( S^\ast=\bigcup_{N\in\mathcal{N}}S[N] \).
If \( S[N]\cap S[N']=\emptyset \) for distinct \( N,N'\in\mathcal{N} \), then the reactions of \( N \) and \( N' \) do not interfere with each other, and \( \bigsqcup\mathcal{N} \) is the \emph{modular composition} of the I/O CRNs in \( \mathcal{N} \).

    \section{Requirements and Robustness}\label{sec:req_and_robust}
This section specifies what a requirement for an input/output chemical reaction network is and what it means for a reaction network to satisfy a requirement robustly.

Intuitively, a requirement for an I/O CRN \( N \) with an initial state \( \xn \) says that, in any context \( \bc=(\bu,V,h) \) satisfying a context assumption \( \alpha(\bc) \), a desired relationship \( \phi(\bu,N_{\bc,\xn}) \) should hold between the input signal \( \bu \) and the output signal \( N_{\bc,\xn} \).
More formally, a \emph{requirement} for \( N \) is an ordered pair \( \Phi=(\alpha,\phi) \),
where the predicates \( \alpha:\;\mathcal{C}_N\rightarrow\{\false,\true\} \) and
\( \phi:\;C[U]\times C[V]\rightarrow\{\false,\true\} \)
are called the \emph{context assumption} and the \emph{input/output requirement} (\emph{I/O requirement}), respectively, of \( \Phi \).
The I/O CRN \( N \) \emph{exactly satisfies} a requirement \( \Phi=(\alpha,\phi) \) with the initial state \( \xn\in\td^S \), and we write \( N,\xn\models\Phi \), if the implication
\begin{equation}\label{eq:exactly_satisfies_requirement}
	\alpha(\bc)\implies\phi(\bu,N_{\bc,\xn})
\end{equation}
holds for every context \( \bc=(\bu,V,h)\in\mathcal{C}_N \).
The I/O CRN \( N \) \emph{exactly satisfies} \( \Phi \), and we write \( N\models\Phi \), if there exists \( \xn\in\td^S \) such that \( N,\xn\models\Phi \).

Two things should be noted about the above definition.
First, a requirement only concerns input and outputs.
Two different I/O CRNs with different sets of state species may satisfy the same requirement.
Second, in order for \( N\models\Phi \) to hold, a \emph{single} initial state \( \xn \) must cause~\eqref{eq:exactly_satisfies_requirement} to hold for \emph{every} context \( \bc \).

It is often sufficient to satisfy a requirement approximately, rather than exactly.
To quantify the approximation here, we use the \emph{supremum norm} defined by \( \norm{f} = \setsup_{t\in\td}|f(t)| \) for all \( f\in C(\td,\mathbb{R}^W) \), where
\[
    |\bx|=\left(\sum_{Y\in W}\bx(Y)^2\right)^{1/2}
\]
is the Euclidean norm on \( \mathbb{R}^W \).
It is well known that \( \norm{f-g} \) is then a well behaved distance between functions \( f,g\in C(\td,\mathbb{R}^W) \), hence also between functions \( f,g\in C[W] \).
For \( f\in C[W] \) and \( \epsilon\in\td \) we thus define the \emph{closed ball of radius} \( \epsilon \) \emph{about} \( f \) \emph{in} \( C[W] \) to be the set
\[
    B_\epsilon(f) = \{g \in C[W] \mid \norm{g-f}\le\epsilon\}.
\]

For \( \epsilon\in\td \) we say that the I/O CRN \( N \) \( \epsilon \)-\emph{satisfies} a requirement \( \Phi=(\alpha,\phi) \) with the initial state \( \xn\in\td^S \), and we write \( N,\xn\models_\epsilon\Phi \),
if the implication
\begin{equation}\label{eq:epsilon_satisfies_requirement}
	\alpha(\bc)\implies(\exists\textbf{v}\in B_\epsilon(N_{\bc,\xn}))\phi(\bu,\textbf{v})
\end{equation}
holds for every context \( \bc=(\bu,h)\in\mathcal{C}_N \).
The I/O CRN \( N \) \( \epsilon \)-\emph{satisfies} \( \Phi \), and we write \( N\models_\epsilon\Phi \), if there exists \( \xn\in\td^S \) such that \( N,\xn\models_\epsilon\Phi \).

It is clear by inspection of~\eqref{eq:exactly_satisfies_requirement} and~\eqref{eq:epsilon_satisfies_requirement} that \( \models \) is equivalent to \( \models_0 \).

We now come to robustness.
Intuitively, an I/O CRN \( N \) with an initial state \( \xn \) robustly \( \epsilon \)-satisfies a requirement \( \Phi=(\alpha,\phi) \) if, for every context \( \bc \) satisfying \( \alpha(\bc) \), the following holds:
For every \qq{\(\bchat \) close to \( \bc \),}
every \qq{\(\xnhat \) close to \( \xn \),}
and every \qq{\(\hat{N} \) close to \( N \),}
the right-hand side of~\eqref{eq:epsilon_satisfies_requirement} holds with \( \hat{N}_{\bchat,\xnhat} \) in place of \( N_{\bc,\xn} \).
To make this intuition precise, we define the three phrases in quotation marks.

We have already used the supremum norm to define the distance \( \norm{f-g} \) between two signals \( f,g\in C[W] \).
We use the same idea and notation to define the distance between two functions \( f,g\in C[W,W'] \) and the closed ball of radius \( \epsilon \) about \( f \) in \( C[W,W'] \).
Given contexts \( \bc=(\bu,V,h) \) and \( \bchat=(\hat{\bu},\hat{V},\hat{h}) \), and given \( \delta_1,\delta_2\in\td \), we say that \( \bchat \) is \( (\delta_1,\delta_2) \)-\emph{close} to \( \bc \) if \( V=\hat{V} \) and \( (\hat{\bu},\hat{h})\in B_{\delta_1}(\bu)\times B_{\delta_2}(h) \).

Given \( \bx,\bxhat\in\td^S \) and \( \delta\in\td \), we say that \( \bxhat \) is \( \delta \)-\emph{close} to \( \bx \) if \( \bxhat\in B_\delta(\bx) \), where the closed ball \( B_\delta(\bx) \) in \( \td^S \) is defined in the obvious way using the Euclidean norm.

The definition of \qq{\(\hat{N} \) close to \( N \)} takes a bit more work, because it allows for the fact that \( \hat{N} \) may be an implementation of \( N \) in which the \qq{rate constants} are only approximately constant.
Nevertheless, the intuition is simple: A \( \delta \)-\emph{perturbation} of \( N \) is a variant \( \hat{N} \) of an I/O CRN in which an \emph{adversary} is allowed to \emph{vary} each rate constant \( k \), subject to the constraint that \( |\hat{k}(t) - k| \leq \delta \) for all \( t \in \td \).

Formally, a \textit{time-dependent reaction} over a finite set \( S \subseteq \mathbf{S} \) is a triple
\[
    \rho=(\br,\bp,\hat{k})\in\mathbb{N}^S\times\mathbb{N}^S\times C(\td, (0,\infty)).
\]
As before, we write \( \br(\rho) = \br \), \( \bp(\rho) = \bp \), and \( \hat{k}(\rho) = \hat{k} \), and we use more intuitive notions like
\[
    X + Z \goesto{\hat{k}} 2Y + Z,
\]
remembering that \( \hat{k} \) is now a function of time, rather than a constant.
An \emph{input/output time-dependent CRN} (\emph{I/O tdCRN}) is then an ordered triple \( \Nhat = (U,\hat{R},S) \), where \( U \) and \( S \) are as in the I/O CRN definition and \( \hat{R} \) is a finite set of time-dependent reactions over \( S \).
The deterministic mass action semantics of an I/O tdCRN \( \Nhat = (U,\hat{R},S) \) is defined in the obvious way, rewriting~\eqref{eq:reaction_rate}--\eqref{eq:mass_action_system_vector} as
\begin{align}
    &\text{rate}_{\bx,\bu}(\rho)(t) = \hat{k}(\rho)(t)(\bx,\bu)^{\mathbf{r}(\rho)}, \\
    &F_Y(\bx,\bu,t) = \sum_{\rho \in \hat{R}} \text{rate}_{\bx,\bu}(\rho)(t)\Delta_\rho(Y), \\
    &y'(t) = F_Y(\bx(t),\bu(t), t), \label{eq:perturbated_ode}\\
    &(\mathcal{E}_Y\mid Y \in S), \\
    &F(\bx,\bu,t) = (F_Y(\bx,\bu,t)\mid Y\in S), \\
    &\bx'(t) = F(\bx(t),\bu(t),t).
\end{align}
The \emph{output signal} \( \Nhat_{\bc,\xn} \) of an I/O tdCRN \( \Nhat \) in the context \( \bc \) with initial state \( \xn \) is defined in the now-obvious manner.

Let \( N = (U,R,S) \) be an I/O CRN, and let \( \delta\in\td \).
A \( \delta \)-\emph{perturbation} of \( N \) is an I/O tdCRN \( \Nhat = (U,\hat{R},S) \) in which \( \hat{R} \) is exactly like \( R \), except that each reaction \( (\br,\bp,k) \) is replaced by a time-dependent reaction \( (\br,\bp,\hat{k}) \) satisfying
\begin{equation}\label{eq:rate_constraint}
    |\hat{k}(t) - k| \leq \delta
\end{equation}
for all \( t\in\td \).

Putting this all together, let \( N=(U,R,S) \) be an I/O CRN, let \( \xn\in\td^S \) be an initial state of \( N \), let \( \Phi=(\alpha,\phi) \) be a requirement for \( N \), let \( \epsilon\in\td \), and let \( \bm{\delta}=(\delta_u,\delta_h,\delta_0,\delta_k)\in(0,\infty)^4 \) be a vector of \emph{strictly positive} real numbers.
We say that \( N \) and \( \xn \) \( \bm{\delta} \)-\emph{robustly} \( \epsilon \)-\emph{satisfy} \( \Phi \), and we write \( N,\xn\models_{\epsilon}^{\bm{\delta}}\Phi \), if, for every \( \bc=(\bu,V,h)\in\mathcal{C}_N \) satisfying \( \alpha(\bc) \), every \( \hat{\bc} \) that is \( (\delta_u,\delta_h) \)-close to \( \bc \), every \( \xnhat \) that is \( \delta_0 \)-close to \( \xn \), and every \( \hat{N} \) that is \( \delta_k \)-close to \( N \), there exists \( \textbf{v}\in B_\epsilon(\hat{N}_{\bchat,\xnhat}) \) such that \( \phi(\bu,\textbf{v}) \) holds.
Finally, we say that \( N \) \( \bm{\delta} \)-\emph{robustly} \( \epsilon \)-\emph{satisfies} \( \Phi \), and we write \( N\models_\epsilon^{\bm{\delta}}\Phi \), if there exists \( \xn\in\td^S \) such that \( N,\xn\models_\epsilon^{\bm{\delta}}\Phi \).

We extend the notations \( \models \), \etc, to the satisfaction of finite sets \( \bm{\Phi} \) of requirements \( \Phi \) in the obvious way.

    \section{Input Enhancement}\label{sec:preproc}
An essential part of our NFA construction is a device that reduces noise in the input signal.
This part of our I/O CRN is a separate module that does not depend on any aspect of the NFA being simulated other than its number of states.
In fact, this preprocessing module consists of several identical submodules, one for each input species of the NFA logic module.
The goal of the preprocessor is to transform the concentration \( x(t) \) of each input species \( X\in U \) into a concentration \( x^*(t) \) that approximates a square-wave.
In particular, when the concentration \( x(t) \) is high, then \( x^*(t) \) is close to \( 1 \), and when \( x(t) \) is low, then \( x^*(t) \) is close to \( 0 \).

We now formally state the requirement of the input enhancer.
Let \( \tau>0 \), and let \( X\in\bS \) be a species.
Define \( \Phi^\supX=\Phi^\supX(\tau)=(\alpha,\phi) \) to be the requirement where the context assumption \( \alpha:C[\{X\}]\rightarrow\{\false,\true\} \) is defined by 
\begin{equation}\label{eq:cleaner_context_assumption}
	\alpha(\bu,V,h)\equiv
		\Big[V=\{X^\ast,\Xbar^\ast\}	\text{ and }	h = h_0\Big],
\end{equation}
where \( h_0 \) is the zero-error projection function from equation~\eqref{eq:zero_error_projection_function}.
Notice that \( \alpha \) requires that the I/O CRN has one input species \( X \) and two output species \( X^\ast \) and \( \Xbar^\ast \).
The two output species are a ``dual rail'' encoding of the input species.
Thus, \( \Xbar^\ast \) represents the Boolean complement of \( X^\ast \) and should be close to \( 0 \) if \( x(t) \) is high and close to \( 1 \) if \( x(t) \) is low.

The I/O requirement \( \phi \) of \( \Phi^\supX \) requires more work to specify, so we begin by defining some helpful terminology and notation.
If \( I=[a,b]\subseteq\td \) is a closed interval, we write \( \text{len}(I) = b - a \) to denote the length of the interval.
If \( I=[a,b] \) and \( \text{len}(I)\ge\tau \), we define the \( \tau \)-\emph{left truncation of} \( I \) to be the subinterval \( I_\tau=[a+\tau,b] \).

Let \( \bu\in C[\{X\}] \) be an input signal, and let \( \bv\in C[\{X^\ast,\Xbar^\ast\}] \) be an output signal.
An \emph{input event} is an ordered pair \( (b,I) \) where \( b\in\{0,1\} \) is a bit, \( I\subseteq\td \) is a closed interval with \( \text{len}(I)\ge\tau \), and \( \bu(t)(X)=b \) for all \( t\in I \).
The set of all input events over \( \bu \) is denoted \( \mathbf{IEV} \).
Intuitively, an input event is a segment of the input signal which has length at least \( \tau \) in which the input is held at \( b \).
An \emph{output event} is an ordered pair \( (b,I) \) where \( b\in\{0,1\} \), \( I\subseteq\td \) is a closed interval, and the following two conditions hold for all \( t\in I \):
\begin{enumerate}
	\item If \( b=1 \), then \( \bv(t)(X^\ast)\ge1 \) and \( \bv(t)(\Xbar^\ast)=0 \).
	\item If \( b=0 \), then \( \bv(t)(X^\ast)=0 \) and \( \bv(t)(\Xbar^\ast)\ge 1 \).
\end{enumerate}
The set of all output events over \( \bv \) is denoted \( \mathbf{OEV} \).

\begin{figure}
	\def\svgwidth{0.49\columnwidth}
	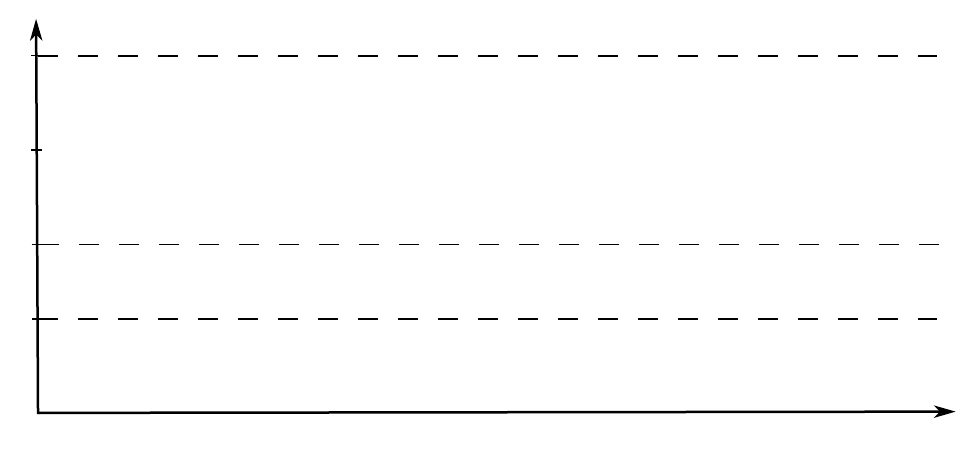
	\def\svgwidth{0.49\columnwidth}
	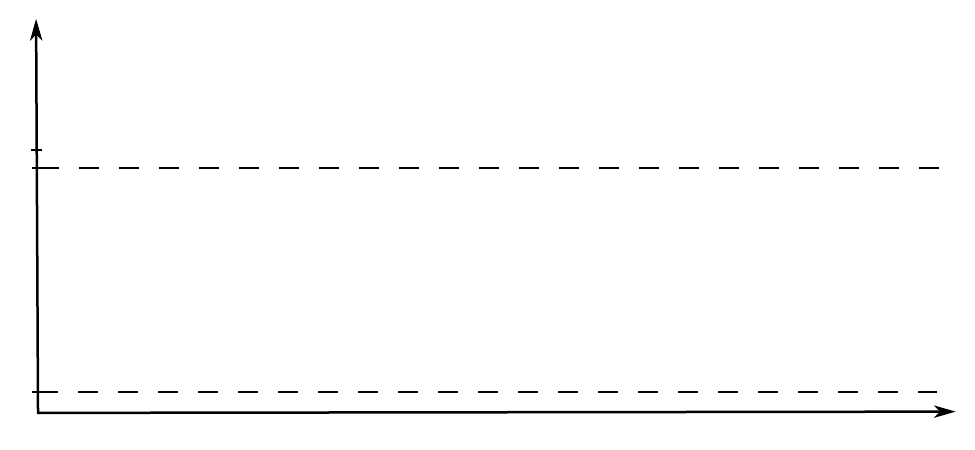
	\caption{\label{fig:cleaner} The two graphs demonstrate the relationship the I/O requirement \( \phi \) imposes on the input and output signals.
	The output signal \( x^\ast(t) \) is essentially an \qq{enhanced} version of the input signal \( x(t) \).}
\end{figure}

We now define the I/O requirement \( \phi \) of \( \Phi^\supX \) to be
\begin{equation}\label{eq:cleaner_io_requirement}
	\phi(\bu,\bv)\equiv
		\big[(b,I)\in\mathbf{IEV}\implies (b,I_\tau)\in\mathbf{OEV}\big].
\end{equation}
Intuitively, \( \Phi \) requires that whenever the input signal has exactly concentration \( b\in\{0,1\} \), the output signal converges to \( b \) and \( 1-b \) in \( \tau \) time.
Therefore the output species encode both the original bit \( b \) and its compliment \( 1-b \).
As an example, suppose an input signal \( x(t) \) contains input events which only get within \( \delta\in(0,\frac{1}{3}) \) of the bits it is encoding.
Then, what we desire is an I/O CRN which is capable of \emph{improving} this signal so that it gets within \( \epsilon<\delta \) of the bits it is encoding and only introducing a delay of at most \( \tau \).
\Cref{fig:cleaner} depicts this relationship in more detail.

We now specify the I/O CRN that is capable of robustly satisfying the requirement \( \Phi^\supX \).
We first state the construction formally and then give an intuitive overview of its operation.
\begin{construction}\label{const:preprocessor}
	Given strictly positive real numbers \( \tau \), \( \epsilon \), and \( \bdelta=(\delta_u,\delta_h,\delta_0,\delta_k) \) where \( \delta_u\in(0,\frac{1}{3}) \), \( \delta_h\in[0,\epsilon) \), and \( \delta_0\in(0,\frac{1}{2}) \), let \( b=\frac{1-\delta_u}{2\delta_u} \) and \( n=\ceil{2\log_b(\frac{8}{\epsilon-\delta_h})} \).
	Define the I/O CRN \( N^\supX=N^\supX(\tau,\epsilon,\bdelta)=(U,R,S) \) where
	\[
		U = \{X\},\qquad
		S = \{X_i\mid 0\le i \le n\} \cup \{X^\ast, \Xbar^\ast\},
	\]
	and where \( R \) consists of the reactions
	\begin{alignat*}{2}
		X + X_i &\goesto{k_1} X + X_{i+1} \quad&&(\forall\; 0 \le i < n)\\
		X_i	&\goesto{k_1} X_0 &&(\forall\; 0 < i \le n)\\
		X_n + \Xbar^\ast &\goesto{k_2} X_n + X^\ast\\
		X^\ast &\goesto{k_2} \Xbar^\ast,
	\end{alignat*}
	and the rate constants \( k_1 \) and \( k_2 \) are defined by
	\begin{align*}
		k_1 &= 
			2\delta_k+\frac{2n\log(2n)}{\tau(1-\delta_u)}+\frac{2}{\tau}\log\left(10\left(\frac{8}{\epsilon-\delta_h}\right)^2\left(\frac{2}{1-\delta_u}\right)^n\right)+\frac{\delta_k(2+\delta_u)}{\delta_u},\\
		k_2 &=\frac{2}{\tau}\log\left(\frac{3}{\epsilon-\delta_h}\right)+4\delta_k.
	\end{align*}

	We also define the initial state \( \xn^\supX \) of \( N^\supX \) by
	\begin{align*}
		\xn^\supX(X^\ast)&=0,\\
		\xn^\supX(\Xbar^\ast)&=1+\delta_0,\\
		\xn^\supX(X_0) &= \frac{10}{\epsilon-\delta_h}\left(\frac{2}{1-\delta_u}\right)^n+\delta_0,\\
		\xn^\supX(X_i) &= 0\quad(\forall\;0<i\le n).
	\end{align*}
\end{construction}

The species of \( N(X) \) are naturally separated into two parts.
The first part is the cascade of species \( X_0,\ldots,X_n \).
This cascade is designed so that every species \( X_i \) \qq{falls down} to \( X_0 \) at a constant rate, and each species \( X_i \) \qq{climbs up} to the next species \( X_{i+1} \) at a rate proportional to the input \( X \).
As a result, whenever the concentration of \( X \) is low, the top of the cascade \( X_n \) is \emph{extremely} low.
Similarly, whenever the concentration of \( X \) is relatively high, the concentration of \( X_n \) becomes relatively high.

The second part of the construction consists of the species \( X^\ast \) and \( \Xbar^\ast \) which are the output species.
The sum of the concentrations of these species is always constant, and the presence of the species \( X_n \) causes \( X^\ast \) to dominate, and the absence of \( X_n \) causes \( \Xbar^\ast \) to dominate.
The cascade and the two species \( X^\ast \) and \( \Xbar^\ast \) collaborate to enhance the input signal.

The length of the cascade, the rate constants, and the initial concentrations are carefully set and depend on the parameters of \( N^\supX \).
For example, the length of the cascade increases as \( \epsilon \) decreases since the output must be enhanced by a larger amount.
The reason for complexity of \( k_1 \), \( k_2 \), \( \xn^\supX(X_0) \) and becomes apparent in the proof of the following theorem.

\begin{theorem}[Input Enhancement Theorem]\label{theorem:preprocessor}
	If \( \tau>0 \), \( \epsilon\in(0,\frac{1}{2}) \), \( \bdelta=(\delta_u,\delta_h,\delta_0,\delta_k) \) with \( \delta_u\in(0,\frac{1}{3}) \), \( \delta_h\in(0,\epsilon) \), \( \delta_0\in(0,\frac{1}{2}) \), \( \delta_k>0 \), and \( N^\supX=N^\supX(\tau,\epsilon,\bdelta) \) and \( \xn^\supX \) are constructed according to Construction~\ref{const:preprocessor}, then
	\begin{equation}
		N^\supX,\xn^\supX\models_\epsilon^{\bdelta}\Phi^\supX(\tau).
	\end{equation}
\end{theorem}

A detailed proof of Theorem~\ref{theorem:preprocessor} is provided in Appendix~\ref{app:preproc}.

    \section{Robust I/O CRN Simulation of NFAs}\label{sec:nfa_sim}
In this section we give the main result of this paper: a uniform translation of an NFA to an I/O CRN that simulates it robustly.
Finite automata are ubiquitous in computer science, but details and notation vary, so we briefly review the specific model used in this paper.
(See, \eg,~\cite{oKoze97}.)

A \emph{nondeterministic finite automaton (NFA)} is an ordered 5-tuple \( M=(Q,\Sigma,\Delta,I,F) \), where
\( Q \) is a finite set of \emph{states};
\( \Sigma \) is a finite \emph{input alphabet};
\( I\subseteq Q \) is the set of \emph{initial states};
\( F\subseteq Q \) is the set of \emph{accepting states};
and \( \Delta:Q\times\Sigma\rightarrow\powerset(Q) \) is the \emph{transition function}.
Here we are using the notation \( \powerset(Q) \) for the \emph{power set} of \( Q \), \ie, the set of all subsets of \( Q \).
When convenient we identify the transition function \( \Delta \) with the set of all \emph{transitions} of \( M \), which are triples \( (q,a,r)\in Q\times\Sigma\times Q \) satisfying \( r\in\Delta(q,a) \).
Informally, the \emph{size} of \( M \) is determined by the three cardinalities \( |Q| \), \( |\Sigma| \), and \( |\Delta| \).

The \emph{extended transition function} of the above NFA \( M \) is the function \( \widehat{\Delta}:\powerset(Q)\times\Sigma^\ast\rightarrow\powerset(Q) \) defined by the recursion
\begin{align*}
	\widehat{\Delta}(A, \lambda) &= A\text{, and}\\
	\widehat{\Delta}(A, wa) &= \bigcup_{q \in \widehat{\Delta}(A, w)}\Delta(q, a)
\end{align*}
for all \( A \subseteq Q \), \( w\in\Sigma^\ast \), and \( a\in\Sigma \), where \( \lambda \) is the \emph{empty string}.
The NFA \( M \) \emph{accepts} an input string \( w \in \Sigma^\ast \) if \( \widehat{\Delta}(I, w) \cap F \neq \emptyset \), \ie, if there is a chain of transitions leading from some state in \( I \) to some state in \( F \).
Otherwise, \( M \) \emph{rejects} \( w \).

Given an NFA \( M=(Q,\Sigma,\Delta,I,F) \), our first objective is to specify a requirement \( \Phi=(\alpha,\phi) \) for an I/O CRN \( N=(U,R,S) \) to simulate \( M \).
The details of \( R \) and \( S \) can be specified later, but \( U \) is an implicit parameter of \( \Phi \), so we at this juncture define the set of input species of \( N \) to be
\begin{equation}\label{eq:input_species}
	U=\{X_a\mid a\in\Sigma\}\cup\{X_r,X_c\},
\end{equation}
where \( r \) (\qq{reset}) and \( c \) (\qq{copy}) are special symbols not occurring in \( \Sigma \).

\begin{figure}
	\centering
	\includegraphics[width=\columnwidth]{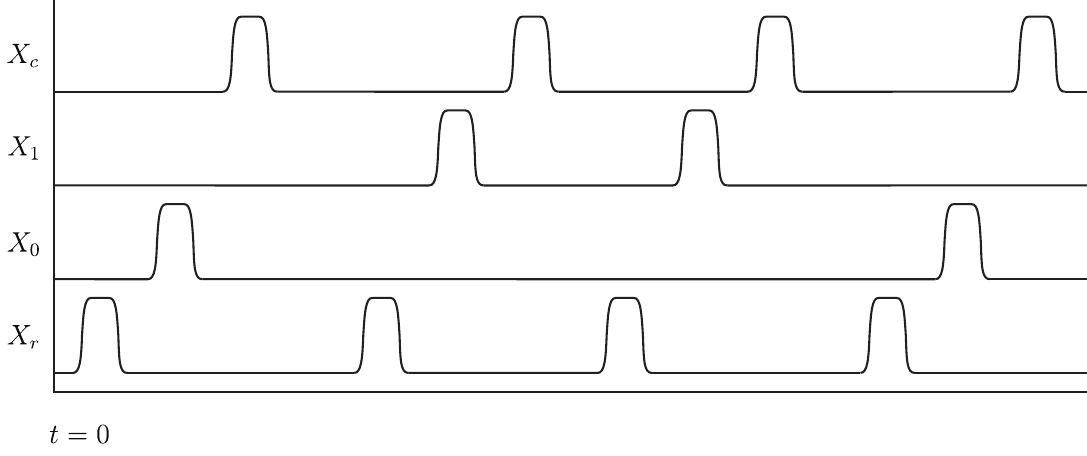}
	\caption{\label{fig:pulse}
		Example input signal for \( 0110 \).}
\end{figure}

We now explain how an input \( w\in\Sigma^\ast \) for \( M \) is provided to \( N \) as a concentration signal.
The intuition is that the input \( w \) is presented as a sequence of pulses in the concentrations of \( |\Sigma|+2 \) species, namely \( \Xr \), \( \Xc \), and \( \Xa \) for each \( a \in \Sigma \).
Each character \( a \) in the string \( w \) is represented by a sequence of three pulses starting with a pulse in the concentration of \( \Xr \), followed by a pulse in the concentration of \( \Xa \), and finally ending with a pulse in the concentration of \( \Xc \).
An example sequence of pulses for the binary string \( 0110 \) is shown in Figure~\ref{fig:pulse}.
To formally specify this intuition as a context assumption, a bit more terminology is needed.
If \( I=[a,b] \) and \( J=[c,d] \) are closed intervals in \( \mathbb{R} \), then \( I \) \emph{lies to the left of} \( J \), and we write \( I<J \), if \( b<c \).

Given an input signal \( \bu\in C[U] \) for \( N \), we define the following.
\begin{enumerate}
    \item
    For \( X\in U \), an \( X \)-\emph{pulse} in \( \bu \) is an interval \( [b,b+4] \), where \( b\in\td \), with the following four properties.
    \begin{enumerate}
        \item For all \( \widehat{X}\in U\setminus\{X\} \) and \( t\in[b,b+4] \), \( \hat{x}(t)=0 \).
        \item For all \( t\in\{b\}\cup[b+3,b+4] \), \( x(t)=0 \).
        \item For all \( t\in[b,b+1]\cup[b+2,b+3] \), \( x(t)\in[0,1] \).
        \item For all \( t\in[b+1,b+2] \), \( x(t)=1 \).
    \end{enumerate}

    \item 
    For \( a\in\Sigma \), an \( a \)-\emph{event} in \( \bu \) is an interval \( [b,b+12] \) such that \( [b,b+4] \) is an \( X_r \)-pulse in \( \bu \), \( [b+4,b+8] \) is an \( X_a \)-pulse in \( \bu \), and \( [b+8,b+12] \) is an \( X_c \)-pulse in \( \bu \).

    \item 
    A \emph{symbol event} in \( \bu \) is an interval \( I\subseteq\td \) that is an \( a \)-event in \( \bu \) for some \( a\in\Sigma \).

    \item
    The input signal \( \bu \) is \emph{proper} if there is a sequence \( (I_i\mid 0\le i < k) \) of symbol events in \( \bu \) such that \( 0\le k\le\infty \), \( 1<I_i<I_{i+1} \) holds for all \( 0\le i < k-1 \), and \( \bu(t)(X)=0 \) holds for all \( X\in U \) and \( t\in\td\setminus\bigcup_{i=0}^{k-1}I_i \).

    \item 
    If \( \bu \) is proper, the sequence \( (I_i\mid 0\le i<k) \) testifies to this fact, \( I_i \) is an \( a_i \)-event for each \( 0\le i<k \), and \( t\in\td \), then the \emph{string presented by} \( \bu \) \emph{at time} \( t \) is the string
    \[
        w(\bu)(t)=a_0a_1\cdots a_{j-1},
    \]
    where \( j \) is the greatest integer such that \( 0\le j<k \) and \( I_j \le t \).

    \item 
    The input signal \( \bu \) is \emph{terminal} if it is proper and the sequence \( (I_i\mid 0\le i < k) \) testifying to this fact is finite, i.e., \( k\in\mathbb{N} \).
    In this case, the \emph{terminus} of \( \bu \) is the time \( \tau(\bu)= \) \emph{if} \( k=0 \) \emph{then} 1 \emph{else} the right endpoint of the interval \( I_{k-1} \), and the \emph{string presented by} \( \bu \) is the string
    \[
        w(\bu)=w(\bu)(\tau(\bu)).
    \]
\end{enumerate}

We now have enough terminology to formally state what it means for an I/O CRN to simulate an NFA.
Given an NFA \( M=(Q,\Sigma,\Delta,I,F) \), we define the requirement \( \Phi=\Phi(M)=(\alpha,\phi) \) as follows.
The context assumption \( \alpha \) of \( \Phi \) is defined by
\begin{equation}\label{eq:req_context_assump}
    \alpha(\bu,V,h)\equiv\;\big[\bu\text{ is terminal}
        \text{ and }V=\{Y_q\mid q\in F\}
        \text{ and }h=h_0\big],
\end{equation}
where \( h_0 \) is the zero-error projection function~\eqref{eq:zero_error_projection_function}.

The I/O requirement \( \phi \) of \( \Phi \) is defined by
\begin{equation}\label{eq:req_io_req}
    \phi(\bu,\bv)\equiv\psi_1\text{ and }\psi_2,
\end{equation}
where \( \psi_1 \) and \( \psi_2 \) are the formulas
\begin{align}
        \psi_1&\equiv\big[M\text{ accepts }w(\bu)\implies(\forall t>\tau(\bu))(\exists Y\in V) \bv(t)(Y)=1\big],\\
        \psi_2&\equiv\big[M\text{ rejects }w(\bu)\implies(\forall t>\tau(\bu))(\forall Y\in V) \bv(t)(Y)=0\big].
\end{align}
The two parts \( \psi_1 \) and \( \psi_2 \) of the I/O requirement correspond to how the I/O CRN should output \qq{accept} and \qq{reject,} respectively.
If the input string presents a string that should be accepted, \( \psi_1 \) requires that the output signal have at least one species \( Y\in V \) that is held at a value of 1 indefinitely.
Similarly, if the input string should be rejected, \( \psi_2 \) requires that the output signal hold all species in \( V \) at a value of 0 indefinitely.

We now specify our translation of an arbitrary NFA into an I/O CRN that simulates it.
The I/O CRN consists of two separate modules: the input enhancement module from \Cref{sec:preproc}, and a module responsible for the NFA logic.
We begin by defining the I/O CRN that computes the logic of the NFA, and later we join this I/O CRN with the input enhancement module.

\begin{construction}\label{const:crn_translation}
	Given an NFA \( M=(Q,\Sigma,\Delta,I,F) \) and strictly positive real numbers \( \epsilon \) and \( \bm{\delta}=(\delta_u,\delta_h,\delta_0,\delta_k) \) satisfying \( \epsilon > \delta_h+\delta_0 \), we define the I/O CRN \( N^\ast=N^\ast(M,\epsilon,\bm{\delta})=(U^\ast,R^\ast,S^\ast) \) as follows.

	The set \( U^\ast \) is the preprocessed equivalent to the species~\eqref{eq:input_species} specified earlier, \ie,
	\[
		U^\ast=\{X^\ast_a\mid a\in\Sigma\}\cup\{X^\ast_r,X^\ast_c\}.
	\]

	The set \( S^\ast \) contains the following three types of species.
	\begin{enumerate}
		\item \emph{State species}.
		For each state \( q\in Q \) there is a species \( Y_q \).
		Intuitively, the concentration of \( Y_q \) is close to 1 in \( N \) when \( M \) could (as permitted by its nondeterminism) be in state \( q \).

		\item \emph{Portal species}.
		For each state \( q\in Q \) there is a species \( Z_q \) that is used as a buffer to facilitate transitions into the state \( q \).

		\item \emph{Dual species}.
		For each state species \( Y_q \) and portal species \( Z_q \), there are species \( \Yqbar \) and \( \Zqbar \).
		We refer to the species \( Y_q \), \( Z_q \) as \emph{basic species} in order to further distinguish them from their duals \( \Ybar_q \), \( \Zbar_q \).
		Intuitively, a dual of a basic species is one that has exactly the opposite operational meaning, \ie, when \( Y_q \) has high concentration, \( \Ybar_q \) has low concentration and vice versa.
	\end{enumerate}
	We define \( S^\ast \) to be the collection of species of these three types, noting that \( |S^\ast|=4|Q| \).

	The reactions of \( N^\ast \) are of four types, designated as follows.
	\begin{enumerate}
		\item \emph{Reset reactions}.
		For each state \( q\in Q \) we have the reaction
        \begin{equation}\label{eq:reset_reaction}
            X^\ast_r + Z_q \goesto{k_1} X^\ast_r + \Zbar_q.
        \end{equation}

		\item \emph{Transition reactions}.
		For each transition \( (q,a,r)\in\Delta \) of \( M \) we have the reaction
        \begin{equation}\label{eq:transition_reaction}
            X^\ast_a + Y_q + \Zbar_r\goesto{k_1} X^\ast_a + Y_q + Z_r.
        \end{equation}

		\item \emph{Copy back reactions}.
        For each state \( q\in Q \) we have the reactions
		\begin{align}
			X^\ast_c + Z_q + \Ybar_q &\goesto{k_2} X^\ast_c + Z_q + Y_q,\label{eq:copy_reaction_1}\\
			X^\ast_c + \Zbar_q + Y_q &\goesto{k_2} X^\ast_c + \Zbar_q + \Ybar_q\label{eq:copy_reaction_2}.
		\end{align}
		\item \emph{State restoration reactions.} For each state \( q\in Q \) we have the reactions
		\begin{align}
			2 \Yq + \Yqbar &\goesto{k_2} 3 \Yq\label{eq:sr_reaction_1}\\
			2 \Yqbar + \Yq &\goesto{k_2} 3 \Yqbar\label{eq:sr_reaction_2}.
		\end{align}
        Note these reactions are an implementation of the termolecular signal restoration algorithm in~\cite{cKlin16}.
	\end{enumerate}
	The rate constants \( k_1 \) and \( k_2 \) are defined by
	\begin{align}
		k_1&=\frac{30|Q|}{\epsilon-\delta_h-\delta_0},\label{eq:k1_specification}\\
        k_2&=18\log\left(\frac{20|Q|}{\epsilon-\delta_h-\delta_0}\right)\label{eq:k2_specification}.
	\end{align}
	We define \( R^\ast \) to be the collections of reactions of these four types, noting that \( |R^\ast|=|\Delta| + 5|Q| \).
	We also note that \( U^\ast\cap S^\ast=\emptyset \) and species in \( U^\ast \) only appear as catalysts in \( R^\ast \), so \( N^\ast \) is indeed an I/O CRN.
\end{construction}

Intuitively, \( N^\ast \) simulates the NFA \( M \) in the following way.
The state species \( Y_q \) and \( \Ybar_q \) for \( q\in Q \) are used to store the states that \( M \) could be in at any time.
More specifically, these species encode the set \( \deltaof{w} \) where \( w \) is the string processed so far.
Whenever the input signal provides another symbol event to \( N^\ast \), it processes the event in three stages, each corresponding to the three pulses of the symbol event.
The first pulse of a symbol event is the \qq{reset} pulse via the species \( X_r^\ast \).
When \( N^\ast \) receives this pulse, it forces all of the concentration of the portal species \( Z_q \) into the species \( \Zbar_q \) using the reactions of equation~\eqref{eq:reset_reaction}.
After the \( X_r^\ast \) pulse is completed, every \( Z_q \) species has concentration close to 0 and every \( \Zqbar \) species has concentration close to 1.
This reset process prepares these portal species to compute the transition function.

The second pulse of the symbol event is an \( X_a^\ast \) pulse for some symbol \( a\in\Sigma \).
When this pulse arrives, \( N^\ast \) computes the transition function of the NFA \( M \) using the reactions from equation~\eqref{eq:transition_reaction}.
Therefore, after this pulse is processed, the portal species \( Z_q \) will be close to 1 if and only if \( q\in\deltaof{wa} \).

The last pulse of the symbol event is the \qq{copy} pulse via the species \( X_c^\ast \).
During this pulse, \( N^\ast \) copies the values of the portal species \( Z_q,\Zbar_q \) back into the state species \( Y_q,\Ybar_q \) using reactions~\eqref{eq:copy_reaction_1} and~\eqref{eq:copy_reaction_2}.
Therefore, after the \( X_c^\ast \) pulse has been processed, the set \( \deltaof{wa} \) is encoded into the state species \( Y_q,\Ybar_q \) which completes the computation.

Finally, the reactions from equations~\eqref{eq:sr_reaction_1} and~\eqref{eq:sr_reaction_2}  ensure that the state will remain valid in the absence of a symbol event indefinitely. 

Two observations concerning \Cref{const:crn_translation} are useful.
First, that \( N^\ast \) is designed to simulate the nondeterminism of \( M \) in real time by computing all transitions in parallel.
Second, this parallelism causes \emph{leak} from one state to the next proportional to the number of states \( |Q| \).
This leak causes the simulation to fail if the input signal is too noisy.
To mitigate this, the input enhancer from \Cref{sec:preproc} preprocesses the noisy input signal in order to present a signal to \( N^\ast \) that guarantees correct simulation.

We now specify the complete I/O CRN that simulates the NFA which includes the input enhancement module.
Recall that our set of input species \( U \) consists of \( |\Sigma|+2 \) elements, one for each symbol in the input alphabet \( \Sigma \) and two for the special symbols \( r \) and \( c \).
Our preprocessing module consists of one input enhancing I/O CRN for each input species.

\begin{construction}\label{const:rbfa_complete_translation}
    Given NFA \( M=(Q,\Sigma,\Delta,I,F) \) and strictly positive real numbers \( \epsilon \) and \( \bdelta=(\delta_u,\delta_h,\delta_0,\delta_k) \), we define the family of I/O CRNs \( \mathcal{N}=\mathcal{N}(M,\epsilon,\bdelta) \) by
    \begin{equation}
        \mathcal{N} = \{N^\ast\}\cup\big\{N^{(X_a)}\mid a\in\Sigma\cup\{r,c\}\big\},
    \end{equation}
    where \( N^\ast=N^\ast(M,\epsilon,\bdelta) \) is constructed according to Construction~\ref{const:crn_translation} and \( N^{(X_a)}=N^{(X_a)}(\frac{1}{2},\gamma,\bdelta^\ast) \) for each \( a\in\Sigma\cup\{r,c\} \) is constructed according to Construction~\ref{const:preprocessor} where
    \begin{equation}
        \gamma=\frac{\epsilon-\delta_h-\delta_0}{(34|Q|)^4}\label{eq:gamma_specification}
    \end{equation}
    and \( \bdelta^\ast=(\delta_u,0,\delta_0,\delta_k) \).

    We also define the I/O CRN \( N=N(M,\epsilon,\bdelta)=(U,R,S) \) to be the join of this family of I/O CRNs
    \begin{equation}
        N = \bigsqcup\mathcal{N}.
    \end{equation}
\end{construction}

Note that \( N \) from Construction~\ref{const:rbfa_complete_translation} is indeed an I/O CRN because  \( \mathcal{N} \) is  \emph{modular}, and the set of input species \( U \) matches that of equation~\eqref{eq:input_species} defined earlier.

We now state the main theorem of this paper.

\begin{theorem}\label{theorem:rbfa}
    If \( M=(Q,\Sigma,\Delta,I,F) \) is an NFA and \( \epsilon \), \( \bm{\delta}=(\delta_u,\delta_h,\delta_0,\delta_k) \) are strictly positive real numbers satisfying
    \begin{align}
        \delta_u,\delta_h,\delta_0,\delta_k&<\frac{1}{20},\\
        \delta_h+\delta_0&<\epsilon,
    \end{align}
    and \( N=N(M,\epsilon,\bdelta) \) is constructed according to Construction~\ref{const:rbfa_complete_translation}, then
    \begin{equation}
        N\models_\epsilon^{\bm{\delta}}\Phi(M).
    \end{equation}
\end{theorem}

The rest of this section is devoted to proving \cref{theorem:rbfa}.
We begin by assuming the hypothesis and construct an initial state for \( N \) that we use to simulate the NFA.
Since \( N \) contains many input enhancement I/O~CRNs, we initialize \( N \) so that each of them are initialized properly according to Construction~\ref{const:preprocessor}.
Therefore, let \( N^\supX=(U^\supX,R^\supX,S^\supX) \) be the input enhancement module of \( N \) for \( X\in U \), and let \( \xn^\supX \) be the initial state of \( N^\supX \) constructed according to Construction~\ref{const:preprocessor}.
Now let \( \xn \) be a state of \( N \) defined by
\begin{alignat*}{2}
    &(\forall q\in I) &&\xn(\Yq)=1=1-\xn(\Yqbar),\\
    &(\forall q\in Q\setminus I)\quad &&\xn(\Yq)=0=1-\xn(\Yqbar),\\
    &(\forall q\in Q) &&\xn(\Zq)=0=1-\xn(\Zqbar),
\end{alignat*}
and
\[
    (\forall X\in U)(\forall \Xhat\in S^{(X)})\quad\xn(\Xhat)=\xn^{(X)}(\Xhat).
\]
The initial state \( \xn \) ensures that every input enhancement module \( N^\supX \) is initialized properly .
The initial state \( \xn \) is also defined so that the state species encode the set of start states \( I \) and the portal species encode the empty set.
We also note that \( \xn(Y_q)+\xn(\Ybar_q)=1 \) and \( \xn(Z_q)+\xn(\Zbar_q)=1 \) for all \( q\in Q \).

Each input enhancement module \( N^\supX = N^\supX(\frac{1}{2},\gamma,\delta^\ast) \) is constructed with delay \( \tau = \frac{1}{2} \) and robustness parameters \( \delta^\ast = (\delta_u,0,\delta_0,\delta_k) \).
A measurement perturbation of \( \delta_h=0 \) is used since the enhanced signals are internally used by \( N^\ast \) and never measured.
The number \( \gamma \) throttles the leak introduced by the state species of \( Q \).
Furthermore, we know that \( N,\xn\models_\gamma^{\bdelta^\ast}\Phi^\supX \) for each \( X\in U \) where \( \Phi^\supX=\Phi^\supX(\frac{1}{2}) \) is the input enhancement requirement from \cref{sec:preproc}.
This follows from \cref{theorem:preprocessor} which says that \( N^\supX,\xn^\supX\models_\gamma^{\bdelta^\ast}\Phi^\supX \) along with the fact that \( N \) is a modular composition I/O CRNs which includes \( N^{(X)} \) for each \( X\in U \).

Since each input enhancer satisfies its requirement, each input event will be enhanced to have at most \( \gamma \) error.
This is an important step for \( N^\ast \) to simulate the NFA because of the leak introduced by the state species.

We now enumerate the ODEs generated by \( N \).
Using the mass action function~\eqref{eq:perturbated_ode}, for each \( q\in Q \), the ODEs of the species \( \Yq,\Zq,\Yqbar,\Zqbar \) of \( N \) are
\begin{align}
    \D{\yq} &= k_2\xcs\zq\yqbar - k_2\xcs\zqbar\yq + k_2\yq^2\yqbar - k_2\yq\yqbar^2,\label{eq:first_ode_eq}\\
    \D{\zq} &= -k_1\xrs\zq + \hspace*{-2ex}\sum_{(s,a,q)\in\Delta}\hspace*{-2ex}k_1\xas\ys\zqbar,\\
    \D{\yqbar}&=-\D{\yq},\\
    \D{\zqbar}&=-\D{\zq},
\end{align}
respectively.

Notice that \( \D{\yq}+\D{\yqbar}=0 \) and \( \D{\zq}+\D{\zqbar}=0 \).
This implies that the sum of the concentrations of \( \Yq \) and \( \Yqbar \) is constant and the sum of the concentrations of \( \Zq \) and \( \Zqbar \) is constant.
Unfortunately, we cannot assume these sums are 1 because of the initial state perturbation.
Therefore, for each \( q\in Q \) we define the constants 
\begin{align}
    \pyq &= \xn(\Yq) + \xn(\Yqbar)\\
    \pzq &= \xn(\Zq) + \xn(\Zqbar),
\end{align}
noting that \( 1-\delta_0 < \pyq, \pzq < 1+\delta_0 \).

We prove that \( N,\xn\models_\epsilon^\bdelta\Phi \) by showing that \( N \) and \( \xn \) robustly satisfy a family of weaker requirements.
To formally state these requirements, more notation and terminology is needed.

For \( A\subseteq Q \) we use \( Y_A=\{\Yq\mid q\in A\} \) and \( Z_A=\{\Zq\mid q\in A\} \) to denote the set of all state species of \( A \) and portal species of \( A \), respectively.
For \( B\subseteq Q \) and vector \( \bx\in\td^{Y_Q} \), we say that \( Y_Q \) \emph{encodes} \( B \) \emph{in} \( \bx \) if \( (\forall q\in B) \) \( \bx(Y_q)=\pyq \) and \( (\forall q\in Q\setminus B) \) \( \bx(Y_q)=0 \).

We also have terminology for approximately encoding a set.
For \( \eta\ge0 \), we say that \( Y_Q \) \( \eta \)-\emph{encodes} \( B \) \emph{in} \( \bx \) if \( (\forall q\in B) \) \( |\pyq-\bx(\Yq)|<\eta \) and \( (\forall q\in Q\setminus B) \) \( \bx(\Yq)<\eta \).
We extend this terminology to the set of portal species \( Z_Q \) in the obvious way.
Furthermore, because an I/O CRN produces a solution of states \( \bx(t) \) that are indexed by time, we occasionally refer to encoding sets \emph{at time} \( t \) when the state \( \bx(t) \) is clear from context.

We now specify the family of requirements.
For \( w\in\Sigma^\ast \), let \( \Phi_w=(\alpha_w,\phi_w) \) be a requirement where \( \alpha_w \) is defined by
\begin{equation}
    \alpha_w(\bu,V,h)\equiv\big[\alpha(\bu,Y_F,h)\text{ and }w(\bu)=w\text{ and }V=Y_Q\big],
\end{equation}
and where \( \phi_w \) is defined by
\begin{equation}
    \phi_w(\bu,\bv)\equiv\,(\forall t\ge\tau(\bu))\big[Y_Q\text{ encodes }\deltaof{w}\text{ in }\bv(t)\big].
\end{equation}
Therefore the requirement \( \Phi_w \) requires that if the I/O CRN receives an input that presents the string \( w\in\Sigma^\ast \), then after processing \( w \) it must output an encoding of \( \deltaof{w} \).

To show that Theorem~\ref{theorem:rbfa} holds, we prove that \( N,\xn\models_\eta^{\bdelta^\ast}\Phi_w \) holds for all \( w\in\Sigma^\ast \) where \( \eta \) is the constant
\begin{equation}\label{eq:eta_specification}
    \eta = \frac{\epsilon-\delta_h-\delta_0}{(80 |Q|)^2}.
\end{equation}
We prove this via induction over the strings \( w\in\Sigma^\ast \) via the following two lemmas, and then show these lemmas suffice to prove \Cref{theorem:rbfa}.

\begin{lemma}[Base Case]\label{lemma:base_case}
   \( N,\xn\models_\eta^{\bdelta^\ast}\Phi_\lambda \).
\end{lemma}

\begin{lemma}[Induction Step]\label{lemma:induction_step}
    For all \( w\in\Sigma^\ast \) and \( a\in\Sigma \) 
    \begin{equation}
        N,\xn\models_\eta^{\bdelta^\ast}\Phi_w
            \implies N,\xn\models_\eta^{\bdelta^\ast}\Phi_{wa}.
    \end{equation}
\end{lemma}

\begin{proof}[Proof of Theorem~\ref{theorem:rbfa}]
    Assume the hypothesis.
    Let \( \bc=(\bu,V,h) \) be a context that satisfies \( \alpha(\bc) \),
    let \( \bchat=(\buhat,V,\hhat) \) be \( (\delta_u,\delta_h) \)-close to \( \bc \),
    let \( \xnhat \) be \( \delta_0 \)-close to \( \xn \),
    let \( \Nhat \) be \( \delta_k \)-close to \( N \),
    let \( w=w(\bu) \), and let \( \bchat_w=(\buhat,Y_Q,h_0) \).
    It suffices to show that \( \Nhat_{\bchat,\xnhat} \) is \( \epsilon \)-close to a signal \( \bv\in C[V] \) such that \( \Phi(\bu,\bv) \) is satisfied.

    By the induction of Lemmas~\ref{lemma:base_case} and~\ref{lemma:induction_step}, we know that \( N,\xn\models_\eta^{\bdelta^\ast}\Phi_{w} \).
    It follows that \( Y_Q \) \( \eta \)-encodes \( \deltaof{w} \) in \( \Nhat_{\bchat_w,\xnhat}(t) \) for all \( t\ge\tau(\bu) \).
    If the NFA \( M \) accepts the string \( w \), then \( F\cap\deltaof{w}\ne\emptyset \), so there exists a \( q\in F \) such that \( \Nhat_{\bchat_w,\xnhat}(t)(Y_q)>\pyq-\eta \) for all \( t\ge\tau(\bu) \).
    Since the perturbed initial state \( \xnhat \) is \( \delta_0 \)-close to \( \xn \), it follows that \( \pyq>1-\delta_0 \).
    Thus,
    \[
       \Nhat_{\bchat_w,\xnhat}(t)(Y_q)>1-\delta_0-\eta 
    \]
    for all \( t\ge\tau(\bu) \).
    We also know that the only difference between \( \Nhat_{\bchat_w,\xnhat}(t) \) and \( \Nhat_{\bchat,\xnhat} \) is the effect of the measurement perturbation by \( \delta_h \).
    Thus, \( \Nhat_{\bchat,\xnhat}(t)(Y_q)>1-\delta_0-\eta-\delta_h \).
    Finally, since \( \epsilon > \delta_h + \delta_0 + \eta \), it follows that \( \Nhat_{\bchat,\xnhat}(t)(Y_q)>1-\epsilon \), and since \( Y_q\in V \), the function \( \Nhat_{\bchat,\xnhat} \) is \( \epsilon \)-close to satisfying \( \psi_1 \) of \( \phi \).

    Similarly, if \( M \) rejects \( w \), then \( F\cap\deltaof{w}=\emptyset \), therefore for all \( Y\in V \) and \( t\ge\tau(\bu) \), \( \Nhat_{\bchat,\xnhat}(t)(Y)<\eta+\delta_h<\epsilon \).
    Therefore \( \Nhat_{\bchat,\xnhat} \) is \( \epsilon \)-close to satisfying \( \psi_2 \) of \( \phi \).
    It follows that \( \Nhat_{\bchat,\xnhat} \) is \( \epsilon \)-close to a function \( \bv\in C[V] \) such that the I/O requirement \( \phi(\bu,\bv) \) holds.
    Therefore \( N,\xn\models_\epsilon^\bdelta\Phi \).
\end{proof}

\begin{figure}[t!]
    \centering
    \includegraphics[height=0.67\textheight]{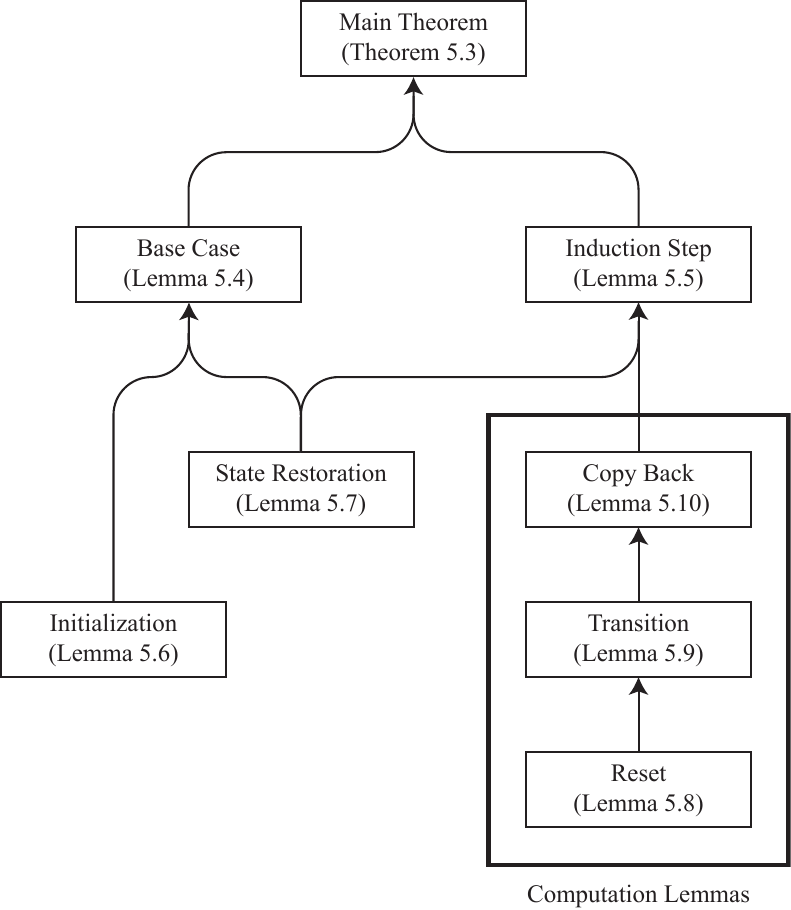}
    \caption{\label{fig:theorem_dependencies}
        Refinements of the main theorem into lemmas}
\end{figure}

It remains to be shown that Lemmas~\ref{lemma:base_case} and~\ref{lemma:induction_step} hold.
The proofs of these are extensive and are broken down into several supporting lemmas which are visualized in Figure~\ref{fig:theorem_dependencies}.

The first two of these supporting lemmas are:

\begin{lemma}[Initialization Lemma]\label{lemma:initialization}
    \( Y_Q \) \( \delta_0 \)-encodes \( I \) at time \( t=\frac{1}{2} \).
\end{lemma}

\begin{lemma}[State Restoration Lemma]\label{lemma:state_restoration}
    Let \( A\subseteq Q \) be a set of states of the NFA, and let \( t_1,t_2\in\td \) be times such that \( t_1+\frac{1}{2}\le t_2 \).
    If the following two conditions hold:
    \begin{enumerate}
        \item
        \( Y_Q \) \( \frac{1}{20} \)-encodes \( A \) at time \( t_1 \) and
         
         \item
         \( x^\ast(t)\le\gamma \) for all \( t\in[t_1,t_2] \) and for each \( X^\ast\in U^\ast \),
    \end{enumerate}
    then \( Y_Q \) \( \eta \)-encodes \( A \) for all \( t\in[t_1 + \frac{1}{2}, t_2] \).
\end{lemma}

Lemma~\ref{lemma:initialization} requires that during the first half-second, the encoding of the initial states is not negatively affected.
This ensures that the input enhancement modules will activate without any noise accumulating in the state species.
The proof of Lemma~\ref{lemma:initialization} is included in Appendix~\ref{app:initialization}.

Lemma~\ref{lemma:state_restoration} states that if the state species of \( N \) are approximately encoding a set of states \( A \) after the last symbol event of its input, then not only will \( N \) continue to encode \( A \), it will \emph{improve} the accuracy of its encoding to \( \eta \).
This lemma serves two purposes:
to \emph{restore} the accuracy of an encoding after processing a symbol event, and to \emph{maintain} that accuracy as long as no more symbol events arrive.
The proof of Lemma~\ref{lemma:state_restoration} is included in Appendix~\ref{app:state_restoration}.

Using the initialization and state restoration lemmas, we can now prove the base case of the induction.

\begin{proof}[Proof of Lemma~\ref{lemma:base_case}]
    Let \( \bc=(\bu,V,h) \) be a context satisfying \( \alpha_\lambda(\bc) \),
    let \( \bchat=(\buhat,V,h) \) be \( (\delta_u,0) \)-close to \( \bc \),
    let \( \xnhat \) be \( \delta_0 \)-close to \( \xn \), and
    let \( \Nhat \) be \( \delta_k \)-close to \( N \).
    To show that \( N,\xn\models_\eta^{\bdelta^\ast}\Phi_\lambda \), we now only need to show that \( \Nhat_{\bchat,\xnhat} \) is \( \eta \)-close to a signal \( \bv\in C[V] \) that satisfies \( \phi_\lambda(\bu,\bv) \).
    Let \( \bxhat(t) \) be the unique solution of the IVP defined by \( \Nhat \) and the initial condition \( \xnhat \).
    It now suffices to show that \( Y_Q \) \( \eta \)-encodes \( I \) for all \( t\ge\tau(\bu) = 1 \).

    By Lemma~\ref{lemma:initialization}, we know that \( Y_Q \) \( \delta_0 \)-encodes \( I \) in state \( \bxhat(\frac{1}{2}) \).
    Then the hypothesis of Lemma~\ref{lemma:state_restoration} is satisfied with \( A=I \) and \( t_1=\frac{1}{2} \).
    Since \( w(\bu)=\lambda \), no symbol event will ever occur in the input.
    It follows that every species \( X\in U \) will have a concentration less than \( \delta_u \) in \( \buhat \) for all \( t\in\td \).
    By Theorem~\ref{theorem:preprocessor}, all the input enhancers will activate by time \( t=\frac{1}{2} \), and so each enhanced signal \( X^\ast\in U^\ast \) will be held at a concentration less than \( \gamma \) for all \( t\ge\frac{1}{2} \).
    Since this will remain true indefinitely, any choice of \( t_2\ge 1 \) will satisfy the hypothesis of Lemma~\ref{lemma:state_restoration}.
    Thus, Lemma~\ref{lemma:state_restoration} tells us \( Y_Q \) will \( \eta \)-encode \( I \) for all \( t\ge 1 \).
\end{proof}

We now turn our attention to proving Lemma~\ref{lemma:induction_step}.
Let \( w\in\Sigma^\ast \) and \( a\in\Sigma \) and assume the inductive hypothesis \( N,\xn\models_\eta^{\bdelta^\ast}\Phi_w \) holds.
Let \( \bc=(\bu,V,h) \) be a context satisfying \( \alpha_{wa}(\bc) \),
let \( \bchat=(\buhat,V,h) \) be \( (\delta_u,0) \)-close to \( \bc \),
let \( \xnhat \) be \( \delta_0 \)-close to \( \xn \), and
let \( \Nhat \) be \( \delta_k \)-close to \( N \).
It suffices to show that \( \Nhat_{\bchat,\xnhat} \) is \( \eta \)-close to a function \( \bv\in C[V] \) which satisfies \( \phi_{wa}(\bu,\bv) \).
We must show that for all \( t\ge \tau(\bu) \) the set \( Y_Q \) \( \eta \)-encodes \( \deltaof{wa} \) at time \( t \).

Let \( I=[b,b+12] \) be the final symbol event of the input \( \bu \).
Then we know that \( I \) is an \( a \)-event and that \( \tau(\bu)=b+12 \).
All the remaining work of proving Lemma~\ref{lemma:induction_step} involves closely examining the behavior of \( \Nhat \) during the the \( a \)-event of \( [b,b+12] \).
Recall that an \( a \)-event consists of three pulses as shown in Figure~\ref{fig:a_event_timing_graph}:

\begin{figure}
    \centering
    \includegraphics{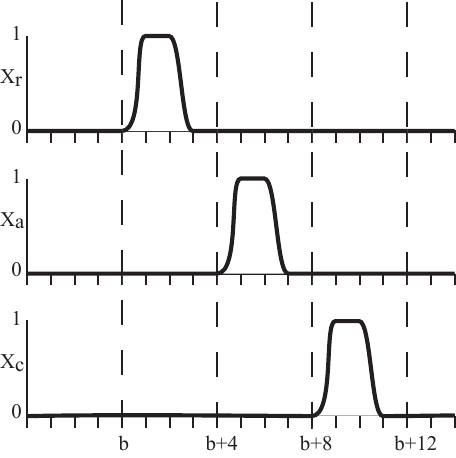}
    \caption{\label{fig:a_event_timing_graph}
        The \( X_r \)-, \( X_a \),- and \( X_c \)-pulses of the final \( a \)-event.
    }
\end{figure}

\begin{enumerate}
    \item an \( X_r \)-pulse during \( [b,b+4] \) that resets portal species \( Z_Q \) to encode \( \emptyset \),
    \item an \( X_a \)-pulse during \( [b+4,b+8] \) that computes the transition function of the NFA and stores the result in the portal species, and
    \item an \( X_c \)-pulse that copies the values of the portal species back into the state species \( Y_Q \).
\end{enumerate}
Since the \( a \)-event is partitioned into three separate pulses, it is natural to break the proof into three parts, each corresponding to one of the pulses.

\begin{lemma}[Reset Lemma]\label{lemma:reset}
    \( Z_Q \) \( \eta \)-encodes \( \emptyset \) at time \( b+4 \).
\end{lemma}

\begin{lemma}[Transition Lemma]\label{lemma:transition}
    \( Z_Q \) \( \frac{1}{20} \)-encodes \( \deltaof{wa} \) during the interval \( [b+8,b+12] \).
\end{lemma}

\begin{lemma}[Copy Back Lemma]\label{lemma:copyback}
    \( Y_Q \) \( \frac{1}{20} \)-encodes \( \deltaof{wa} \) at time \( b+11.5 \).
\end{lemma}

These three lemmas are the \emph{computation lemmas} and their proofs are provided in Appendix~\ref{app:computation_lemmas}.
Intuitively, the Reset Lemma simply says that the portal species are ``reset'' to encode the empty set during the \( X_r \)-pulse so that the transition function of the NFA can be properly computed during the \( X_a \)-pulse.
Similarly, the Transition Lemma says that the transition function is successfully computed during the \( X_a \)-pulse and maintained through the \( X_c \)-pulse.
Finally, the Copy Back Lemma says that near the end of the \( X_c \)-pulse, \( Y_Q \) is properly encoding the correct set of states \( \deltaof{wa} \).

Using the above computation lemmas, we now finish the proof of the induction step. 

\begin{proof}[Proof of Lemma~\ref{lemma:induction_step}]
    By Lemmas~\ref{lemma:reset},~\ref{lemma:transition}, and~\ref{lemma:copyback}, we know that \( Y_Q \) \( \frac{1}{20} \)-encodes the set \( \deltaof{wa} \) at time \( b+11.5 \).
    Since this \( a \)-event \( [b,b+12] \) is the last event in the input signal \( \bu \), the terminus of \( \bu \) is \( \tau(\bu)=b+12 \).
    This means that \( x(t) < \delta_0 \) for each \( X\in U \) and for all time \( t\ge b+11 \).
    By Theorem~\ref{theorem:preprocessor}, the input enhancers will clean up the input signals so that \( x^\ast(t)<\gamma \) for all \( t\ge b+11.5 \).
    Finally, by Lemma~\ref{lemma:state_restoration}, we know that \( Y_Q \) will \( \eta \)-encode \( \deltaof{wa} \) for all \( t\ge b+12 = \tau(\bu) \).
    Thus, \( \Nhat_{\bchat,\buhat} \) is \( \eta \)-close to a signal \( \bv\in C[V] \) that satisfies the I/O requirement \( \phi_{wa} \).
\end{proof}

This concludes the proofs of the main supporting lemmas of Theorem~\ref{theorem:rbfa}.
Proofs of Lemmas~\ref{lemma:initialization} and~\ref{lemma:state_restoration} are are provided in Appendix~\ref{app:initialization} and~\ref{app:state_restoration}, respectfully, and the proofs of the computation lemmas are given in Appendix~\ref{app:computation_lemmas}.

    \section{Conclusion}\label{sec:conclusion}
Unlike traditional CRNs where input is constrained to an initial state, I/O CRNs have designated input species for providing input signals over time.
These input species may only be used catalytically which requires the I/O CRN to access its input non-destructively.
We also introduced a notion of satisfying a requirement \emph{robustly}. 
In particular, robust I/O CRNs must satisfy their requirement even in the presence of adversarial perturbations to their input signals, output measurement (decision), initial concentrations, and reaction rate constants.

Using these definitions, we showed that any non-deterministic finite automaton can be translated into an I/O CRN that robustly simulates it.
Our translation also \emph{efficiently} simulates an NFA by exploiting the inherent parallelism in the I/O CRN model.
Specifically, the non-determinism (existential quantification) is achieved directly by simulating all possible paths through the finite-state machine in parallel.

A key contribution of this paper is our \emph{proof} that the translation robustly simulates the NFA in an adversarial environment.
The proof was refined into two main parts corresponding to the two modules of our construction.
The first module diminishes noise in the input signals to an acceptable level and reshapes it to be closer to a square wave.
We intentionally specified this module separately so that it can be used in other molecular computation devices.
The second module is responsible for computing the NFA transition function and maintains its state until the next symbol event occurs.

We hope that future research will improve our robustness bounds and shed new light on possible tradeoffs among the number of species, the computation time, and robustness. 

    \subsubsection*{Acknowledgments}
We thank students and faculty auditors in our spring, 2015, molecular programming course for helpful and challenging discussions.
We thank Neil Lutz for a discussion that helped us finalize the I/O CRN model.

    \bibliographystyle{plain}

\begin{thebibliography}{10}

\bibitem{jAlze97}
Horst Alzer.
\newblock On some inequalities for the incomplete gamma function.
\newblock {\em Mathematics of Computation}, 66(218):771--778, 1997.

\bibitem{jAnAsEi08a}
Dana Angluin, James Aspnes, and David Eisenstat.
\newblock A simple population protocol for fast robust approximate majority.
\newblock {\em Distributed Computing}, 21(2):87--102, 2008.

\bibitem{jAAER07}
Dana Angluin, James Aspnes, David Eisenstat, and Eric Ruppert.
\newblock The computational power of population protocols.
\newblock {\em Distributed Computing}, 20(4):279--304, 2007.

\bibitem{oApos69}
Tom~M Apostol.
\newblock {\em Calculus, Volume 2: Multi-Variable Calculus and Linear Algebra
  with Applications to Differential Equations and Probability}.
\newblock John Wiley \& Sons, 1969.

\bibitem{jAris65}
Rutherford Aris.
\newblock Prolegomena to the rational analysis of systems of chemical
  reactions.
\newblock {\em Archive for Rational Mechanics and Analysis}, 19(2):81--99,
  1965.

\bibitem{oAstMur08}
Karl~Johan Astr{\"o}m and Richard~M Murray.
\newblock {\em Feedback Systems: An Introduction for Scientists and Engineers}.
\newblock Princeton University Press, 2008.

\bibitem{cBSJDTW17}
Stefan Badelt, Seung~Woo Shin, Robert~F. Johnson, Qing Dong, Chris Thachuk, and
  Erik Winfree.
\newblock A general-purpose {CRN}-to-{DSD} compiler with formal verification,
  optimization, and simulation capabilities.
\newblock In {\em Proceedings of the 23rd International Conference on {DNA}
  Computing and Molecular Programming}, Lecture Notes in Computer Science,
  pages 232--248, 2017.

\bibitem{cBonPou13}
Filippo Bonchi and Damien Pous.
\newblock Checking {NFA} equivalence with bisimulations up to congruence.
\newblock In {\em Proceedings of the 40th Symposium on Principles of
  Programming Languages}, pages 457--468. ACM, 2013.

\bibitem{jBonPou15}
Filippo Bonchi and Damien Pous.
\newblock Hacking nondeterminism with induction and coinduction.
\newblock {\em Communications of the ACM}, 58(2):87--95, 2015.

\bibitem{jCard14}
Luca Cardelli.
\newblock Morphisms of reaction networks that couple structure to function.
\newblock {\em BMC Systems Biology}, 8:84, 2014.

\bibitem{jCarCsi12}
Luca Cardelli and Attila Csik{\'a}sz-Nagy.
\newblock The cell cycle switch computes approximate majority.
\newblock {\em Scientific Reports}, 2, 2012.

\bibitem{jCDSPCS13}
Yuan-Jyue Chen, Neil Dalchau, Niranjan Srinivas, Andrew Phillips, Luca
  Cardelli, David Soloveichik, and Georg Seelig.
\newblock Programmable chemical controllers made from {DNA}.
\newblock {\em Nature Nanotechnology}, 8(10):755--762, 2013.

\bibitem{jCSHELM12}
Juan Cheng, Sarangapani Sreelatha, Ruizheng Hou, Artem Efremov, Ruchuan Liu,
  Johan R.~C. van~der Maarel, and Zhisong Wang.
\newblock Bipedal nanowalker by pure physical mechanisms.
\newblock {\em Physical Review Letters}, 109:238104, 2012.

\bibitem{oCSWB09}
Matthew Cook, David Soloveichik, Erik Winfree, and Jehoshua Bruck.
\newblock Programmability of chemical reaction networks.
\newblock In Anne Condon, David Harel, Joost~N. Kok, Arto Salomaa, and Erik
  Winfree, editors, {\em Algorithmic Bioprocesses}, Natural Computing Series,
  pages 543--584. Springer, 2009.

\bibitem{cDKTT13}
Frits Dannenberg, Marta Kwiatkowska, Chris Thachuk, and Andrew~J. Turberfield.
\newblock {DNA} walker circuits: Computational potential, design, and
  verification.
\newblock In {\em Proceedings of the 19th International Conference on {DNA}
  Computing and Molecular Programming}, volume 8141 of {\em Lecture Notes in
  Computer Science}, pages 31--45. Springer, 2013.

\bibitem{oDelMur14}
Domitilla Del~Vecchio and Richard~M Murray.
\newblock {\em Biomolecular Feedback Systems}.
\newblock Princeton University Press, 2014.

\bibitem{cDLPSSW12}
David Doty, Jack~H Lutz, Matthew~J Patitz, Robert~T Schweller, Scott~M Summers,
  and Damien Woods.
\newblock The tile assembly model is intrinsically universal.
\newblock In {\em Proceedings of the 53rd Symposium on Foundations of Computer
  Science}, pages 302--310. IEEE, 2012.

\bibitem{jDoBaCh12}
Shawn~M. Douglas, Ido Bachelet, and George~M. Church.
\newblock A logic-gated nanorobot for targeted transport of molecular payloads.
\newblock {\em Science}, 335(6070):831--834, 2012.

\bibitem{jDMTVCS09}
Shawn~M. Douglas, Adam~H. Marblestone, Surat Teerapittayanon, Alejandro
  Vazquez, George~M. Church, and William~M. Shih.
\newblock Rapid prototyping of 3{D} {DNA}-origami shapes with ca{DNA}no.
\newblock {\em Nucleic Acids Research}, pages 1--6, 2009.

\bibitem{oEpsPoj98}
Irving~Robert Epstein and John~Anthony Pojman.
\newblock {\em An Introduction to Nonlinear Chemical Dynamics: Oscillations,
  Waves, Patterns, and Chaos}.
\newblock Oxford University Press, 1998.

\bibitem{oErdTot89}
P{\'e}ter {\'E}rdi and J\'{a}nos T\'{o}th.
\newblock {\em Mathematical Models of Chemical Reactions: Theory and
  Applications of Deterministic and Stochastic Models}.
\newblock Manchester University Press, 1989.

\bibitem{cFLBP17}
Fran{\c{c}}ois Fages, Guillaume Le~Guludec, Olivier Bournez, and Amaury Pouly.
\newblock Strong {T}uring completeness of continuous chemical reaction networks
  and compilation of mixed analog-digital programs.
\newblock In {\em Proceedings of the 15th International Conference on
  Computational Methods in Systems Biology}, pages 108--127. Springer
  International Publishing, 2017.

\bibitem{oGaut98}
Walter Gautschi.
\newblock The incomplete gamma functions since {Tricomi}.
\newblock In Francesco~Giacomo Tricomi, editor, {\em Tricomi's Ideas and
  Contemporary Applied Mathematics}, volume 147, pages 203--237. Accademia
  Nazionale dei Lincei, 1998.

\bibitem{oGuna03}
Jeremy Gunawardena.
\newblock Chemical reaction network theory for in-silico biologists, 2003.
\newblock \url{http://www.jeremy-gunawardena.com/papers/crnt.pdf}.

\bibitem{jHPNDLY11}
Dongran Han, Suchetan Pal, Jeanette Nangreave, Zhengtao Deng, Yan Liu, and Hao
  Yan.
\newblock {DNA} origami with complex curvatures in three-dimensional space.
\newblock {\em Science}, 332(6027):342--346, 2011.

\bibitem{jHenRas15}
Thomas~A Henzinger and Jean-Fran{\c{c}}ois Raskin.
\newblock The equivalence problem for finite automata: technical perspective.
\newblock {\em Communications of the ACM}, 58(2):86--86, 2015.

\bibitem{jKOSY12}
Yonggang Ke, Luvena~L. Ong, William~M. Shih, and Peng Yin.
\newblock Three-dimensional structures self-assembled from {DNA} bricks.
\newblock {\em Science}, 338(6111):1177--1183, 2012.

\bibitem{cKlin16}
Titus~H. Klinge.
\newblock Robust signal restoration in chemical reaction networks.
\newblock In {\em Proceedings of the 3rd International Conference on Nanoscale
  Computing and Communication}, pages 6:1--6:6. ACM, 2016.

\bibitem{oKoze97}
Dexter Kozen.
\newblock {\em Automata and Computability}.
\newblock Springer, 1997.

\bibitem{oLent15}
G{\'a}bor Lente.
\newblock {\em Deterministic Kinetics in Chemistry and Systems Biology: The
  Dynamics of Complex Reaction Networks}.
\newblock Springer, 2015.

\bibitem{jQiaWin11a}
Lulu Qian and Erik Winfree.
\newblock Scaling up digital circuit computation with {DNA} strand displacement
  cascades.
\newblock {\em Science}, 332(6034):1196--1201, 2011.

\bibitem{jQiaWin11}
Lulu Qian and Erik Winfree.
\newblock A simple {DNA} gate motif for synthesizing large-scale circuits.
\newblock {\em Journal of the Royal Society Interface}, 8(62):1281--1297, 2011.

\bibitem{jQiWiBr11}
Lulu Qian, Erik Winfree, and Jehoshua Bruck.
\newblock Neural network computation with {DNA} strand displacement cascades.
\newblock {\em Nature}, 475(7356):368--372, 2011.

\bibitem{jRabSco59}
Michael~O Rabin and Dana Scott.
\newblock Finite automata and their decision problems.
\newblock {\em IBM Journal of Research and Development}, 3(2):114--125, 1959.

\bibitem{jRoth06}
Paul W.~K. Rothemund.
\newblock Folding {DNA} to create nanoscale shapes and patterns.
\newblock {\em Nature}, 440(7082):297--302, 2006.

\bibitem{jSeem82}
Nadrian~C. Seeman.
\newblock Nucleic acid junctions and lattices.
\newblock {\em Journal of Theoretical Biology}, 99(2):237 -- 247, 1982.

\bibitem{jSmit10}
Lloyd~M. Smith.
\newblock Nanotechnology: Molecular robots on the move.
\newblock {\em Nature}, 465(7295):167--168, 2010.

\bibitem{jSCWB08}
David Soloveichik, Matthew Cook, Erik Winfree, and Jehoshua Bruck.
\newblock Computation with finite stochastic chemical reaction networks.
\newblock {\em Natural Computing}, 7(4):615--633, 2008.

\bibitem{jSoSeWi10}
David Soloveichik, Georg Seelig, and Erik Winfree.
\newblock {DNA} as a universal substrate for chemical kinetics.
\newblock {\em Proceedings of the National Academy of Sciences},
  107(12):5393--5398, 2010.

\bibitem{oTesc12}
Gerald Teschl.
\newblock {\em Ordinary Differential Equations and Dynamical Systems}.
\newblock American Mathematical Society, 2012.

\bibitem{jWYEWT11}
Anthony~S. Walsh, Hai{F}ang Yin, Christoph~M. Erben, Matthew J.~A. Wood, and
  Andrew~J. Turberfield.
\newblock {DNA} cage delivery to mammalian cells.
\newblock {\em ACS Nano}, 5(7):5427--5432, 2011.

\bibitem{jWeDaYi12}
Bryan Wei, Mingjie Dai, and Peng Yin.
\newblock Complex shapes self-assembled from single-stranded {DNA} tiles.
\newblock {\em Nature}, 485(7400):623--626, 2012.

\bibitem{oWinf98}
Erik Winfree.
\newblock {\em Algorithmic self-assembly of {DNA}}.
\newblock PhD thesis, California Institute of Technology, 1998.

\bibitem{jWood15}
Damien Woods.
\newblock Intrinsic universality and the computational power of self-assembly.
\newblock {\em Philosophical Transactions of the Royal Society A}, 373(2046),
  2015.

\bibitem{jYTMSN00}
Bernard Yurke, Andrew~J. Turberfield, Allen~P. Mills, Friedrich~C. Simmel, and
  Jennifer~L. Neumann.
\newblock A {DNA}-fuelled molecular machine made of {DNA}.
\newblock {\em Nature}, 406(6796):605--608, 2000.

\bibitem{jZhaSee11}
David~Yu Zhang and Georg Seelig.
\newblock Dynamic {DNA} nanotechnology using strand-displacement reactions.
\newblock {\em Nature Chemistry}, 3(2):103--113, 2011.

\end{thebibliography}

    \appendix
    \newpage

    \section{Proof of Input Enhancement Theorem}\label{app:preproc}
This section is dedicated to proving Theorem~\ref{theorem:preprocessor}.
The proof is naturally partitioned into two parts:
\Cref{sub:cleaner_analysis} is dedicated to the analysis of the ODEs generated by the cascade of species in Construction~\ref{const:preprocessor}, and \cref{sub:cleaner_proof} presents a complete proof of the theorem.

\subsection{Cascade Analysis}\label{sub:cleaner_analysis}
In this section, we only concern ourselves with analyzing systems of ODEs.
The construction below is a simplified specification of the ODEs generated by the cascade from Construction~\ref{const:preprocessor}.
We use \( f \) (\qq{forward}) and \( b \) (\qq{backward}) for the rate constants of climbing up the cascade and falling to the bottom of the cascade, respectively.
Notice that we also fold in the concentration of the input species \( X \) into the constant \( f \).
This simplification allows us to thoroughly analyze the behavior of the cascade whenever \( X \) is held constant and is enough to prove the theorem in the following section.

\begin{construction}\label{const:cascade}
    Given \( f>0 \), \( b>0 \), and \( n\in\mathbb{N} \), let \( x_0,\ldots,x_n:\td\rightarrow\td \) be functions that satisfy the ODEs
    \begin{align}
        \D{x_0} &= \sum_{i=1}^n b x_i - f x_0,\label{eq:cascade_ode_1}\\
        \D{x_i} &= fx_{i-1} - (f+b) x_i\quad\text{for }0<i<n,\label{eq:cascade_ode_2}\\
        \D{x_n} &= fx_{n-1} - b x_n\label{eq:cascade_ode_3}.
    \end{align}
\end{construction}

We will now solve for explicit solutions to an IVP generated by the ODEs above using induction.
These solutions have similar structure, so we define the following family of functions to describe their solution.

\begin{construction}\label{const:ladder_function}
    Given \( f>0 \), \( b>0 \), \( p>0 \) and \( i\in\mathbb{N} \), let \( F_i:\td\rightarrow\td \) be the function
    \begin{equation}\label{eq:ladder_function}
        F_i(t) = p\left(\frac{f}{f+b}\right)^ie^{-(f+b)t}\sum_{k=i}^{\infty}\frac{t^k(f+b)^k}{k!}.
    \end{equation}
\end{construction}

\begin{observation}\label{obs:ladder_function_integral}
    If \( f>0 \), \( b>0 \), \( p>0 \), \( i\in\mathbb{N} \) and \( F_i \) is constructed according to Construction~\ref{const:ladder_function}, then
    \begin{equation}\label{eq:ladder_int}
        \int e^{(f+b)t}\cdot F_i(t)dt = \frac{1}{f}e^{(f+b)t}F_{i+1}(t)+C
    \end{equation}
    for some \( C\in\mathbb{R} \).
\end{observation}
\begin{proof}
    Assume the hypothesis.
    Then by the definition of \( F_i \) from equation~\eqref{eq:ladder_function},
    \begin{align}\label{eq:int1}
        \int e^{(f+b)t}\cdot F_i(t)dt
            &= p\left(\frac{f}{f+b}\right)^i\int\sum_{k=i}^{\infty}\frac{t^k(f+b)^k}{k!}dt.
    \end{align}
    The integral can be evaluated to obtain
    \begin{align*}
        \int\sum_{k=i}^{\infty}\frac{t^k(f+b)^k}{k!}dt
             &= \frac{1}{f+b}\sum_{k=i+1}^{\infty}\frac{t^k(f+b)^k}{k!}+C_1,
    \end{align*}
    for some \( C_1\in\mathbb{R} \).  Inserting this into (\ref{eq:int1}), we obtain
    \begin{align*}
        \int e^{(f+b)t}\cdot F_i(t)dt
            &= p\left(\frac{f}{f+b}\right)^i\left(\frac{1}{f+b}\sum_{k=i+1}^{\infty}\frac{t^k(f+b)^k}{k!}+C_1\right)\\
            &= \frac{1}{f}e^{(f+b)t}F_{i+1}(t)+C
    \end{align*}
    for some \( C\in\mathbb{R} \), which is the right side of (\ref{eq:ladder_int}).
\end{proof}

\begin{lemma}\label{lemma:lower_cascade}
    If \( f>0 \), \( b>0 \), \( p>0 \), \( n\in\mathbb{N} \), and for \( 0\le i\le n \) the functions \( x_i \) and \( F_i \) are constructed according to Construction~\ref{const:cascade} and~\ref{const:ladder_function} such that \( x_0(0)=p \) and \( x_i(0)=0 \) for all \( 0<i\le n \), then for all \( t\in\td \),
    \begin{align}
        x_i(t) &= F_i(t) - F_{i+1}(t)\quad\text{for }0\le i<n,\label{eq:lower_cascade_1}\\
        x_n(t) &= F_n(t).\label{eq:lower_cascade_2}
    \end{align}
\end{lemma}
\begin{proof}
    Assume the hypothesis.
    We begin by proving~\eqref{eq:lower_cascade_1} by induction on \( i \).

    Since \( \sum_{i=0}^n\D{x_i}=0 \), it follows that \( \sum_{i=0}^n x_i(t)=p \) for all \( t\in\td \).
    Therefore~\eqref{eq:cascade_ode_1} can be simplified to
    \[
        \D{x_0} = b (p-x_0) - f x_0,
    \]
    and has solution
    \begin{align*}
        x_0(t) 
            &= p - p\left(\frac{f}{f+b}\right)\left(1-e^{-(f+b)t}\right)
            = F_0(t) - F_1(t).
    \end{align*}

    For the induction step, assume that \( x_i(t)=F_i(t)-F_{i+1}(t) \) for some \( 0\le i<n-1 \).
    By~\eqref{eq:cascade_ode_2}, the derivative of \( x_{i+1} \) is
    \[
        \D{x_{i+1}} = fx_i - (f+b) x_{i+1}.
    \]
    which can be solved using the integrating factor method.
    Thus, we obtain the family of solutions
    \begin{align*}
        x_{i+1}(t)
            &= e^{-(f+b)t}\int e^{(f+b)t}fx_i(t)dt\\
            &= e^{-(f+b)t}\int e^{(f+b)t}f\left(F_i(t)-F_{i+1}(t)\right)dt.
    \end{align*}
    It immediately follows from Observation~\ref{obs:ladder_function_integral} that
    \[
        x_{i+1}(t) = F_{i+1}(t)-F_{i+2}(t) + C\cdot e^{-(f+b)t}
    \]
    for some \( C\in\mathbb{R} \).
    By the initial condition, \( x_{i+1}(0)=0 \), and therefore \( C=0 \).
    This completes the induction and shows that~\eqref{eq:lower_cascade_1} holds.

    It remains to be shown that~\eqref{eq:lower_cascade_2} holds.
    Since \( \sum_{i=0}^n x_i(t)=p \) for all \( t\in\td \), we know that
    \[
        x_n(t) = p - \sum_{i=0}^{n-1} x_i(t)
    \]
    for all \( t\in\td \) which can be written in terms of equation~\eqref{eq:ladder_function} in the following way:
    \[
    	 x_n(t) = F_0(t) + \sum_{i=0}^{n-1} F_{i+1}(t) - \sum_{i=0}^{n-1} F_i(t).
    \]
    Finally, we obtain (\ref{eq:lower_cascade_2}) after canceling terms in the above equation.
\end{proof}

\begin{lemma}\label{lemma:lower_cascade_bound}
    Under the assumptions of Lemma~\ref{lemma:lower_cascade},
    \begin{equation}\label{eq:lower_cascade_bound}
        x_n(t) = F_n(t) > p\left(\frac{f}{f+b}\right)^n\left(1-ne^{-\frac{1}{n}(f+b)t}\right)
    \end{equation}
    for all \( t\in\td \).
\end{lemma}
\begin{proof}
    It suffices to show that
    \begin{equation}\label{eq:gamma_inequality}
        e^{-(f+b)t}\sum_{k=n}^{\infty}\frac{t^k(f+b)^k}{k!} > 1-ne^{-\frac{1}{n}(f+b)t}.
    \end{equation}
    The left-hand side of the~\eqref{eq:gamma_inequality} is related to the incomplete gamma function  \( \gamma \) by
    \[
        e^{-(f+b)t}\sum_{k=n}^{\infty}\frac{t^k(f+b)^k}{k!} = \frac{\gamma(n,(f+b)t)}{(n-1)!}.
    \]
    The incomplete gamma function is well understood and many useful bounds exist.
    One particularly useful bound by Alzer~\cite{jAlze97,oGaut98} is
    \[
        \frac{\gamma(a,x)}{\Gamma(a)} > \left(1-e^{-s_ax}\right)^a,
    \]
    for \( a\ge 1 \) where \( s_a=|\Gamma(1+a)|^{-\frac{1}{a}} \).
    It follows that
    \[
        \frac{\gamma(n,(f+b)t)}{(n-1)!} > \left(1-e^{-s_n(f+b)t}\right)^n,
    \]
    where \( s_n=(n!)^{-\frac{1}{n}} \).
    Since \( n^n\ge n! \), we know \( s_n>\frac{1}{n} \), whence
    \begin{align*}
        \frac{\gamma(n,(f+b)t)}{(n-1)!}
            &> \left(1-e^{-\frac{1}{n}(f+b)t}\right)^n
            > 1-ne^{-\frac{1}{n}(f+b)t}.\qedhere
    \end{align*}
\end{proof}

\begin{corollary}\label{corollary:lower_cascade_bound}
    Under the assumptions of Lemma~\ref{lemma:lower_cascade},
    \begin{equation}\label{eq:lower_cascade_bound_simple}
        x_n(t) = F_n(t) > \frac{p}{2}\left(\frac{f}{f+b}\right)^n
    \end{equation}
    for all \( t\ge \frac{n\log(2n)}{f+b} \).
\end{corollary}

\begin{lemma}\label{lemma:upper_cascade}
    If \( f>0 \), \( b>0 \), \( p>0 \), \( n\in\mathbb{N} \), and for \( 0\le i\le n \) the functions \( x_i \) and \( F_i \) are constructed according to Construction~\ref{const:cascade} and~\ref{const:ladder_function} such that \( x_i(0)=0 \) for all \( 0\le i<n \) and \( x_n(0)=p \), then for all \( t\in\td \),
    \begin{equation}
        x_i(t) = \frac{b}{f}F_{i+1}(t)\quad\text{for }0\le i<n\label{eq:upper_cascade_1}.
    \end{equation}
\end{lemma}
\begin{proof}
    Assume the hypothesis.
    We prove~\eqref{eq:upper_cascade_1} by induction on \( i \).

    Since \( \sum_{i=0}^n\D{x_i}=0 \), it follows that \( \sum_{i=0}^n x_i(t)=p \) for all \( t\in\td \).
    Therefore~\eqref{eq:cascade_ode_1} can be simplified to
    \[
        \D{x_0} = b (p-x_0) - f x_0,
    \]
    and has solution
    \begin{align*}
        x_0(t) 
            = p\left(\frac{b}{f+b}\right)\left(1-e^{-(f+b)t}\right)
            = \frac{b}{f}\cdot F_1(t).
    \end{align*}

    For the induction step, assume that \( x_i(t)=\frac{b}{f}F_{i+1}(t) \) for some \( 0\le i<n-1 \).
    By~\eqref{eq:cascade_ode_2}, the derivative of \( x_{i+1} \) is
    \[
        \D{x_{i+1}} = fx_i - (f+b) x_{i+1}.
    \]
    By the integrating factor method, we obtain the solution
    \[
        x_{i+1}(t)
            = e^{-(f+b)t}\int e^{(f+b)t}fx_i(t)dt
            = e^{-(f+b)t}\int e^{(f+b)t}bF_{i+1}(t)dt.
    \]
    It follows from Observation~\ref{obs:ladder_function_integral} that
    \[
        x_{i+1}(t) = \frac{b}{f}F_{i+2}(t) + Ce^{-(f+b)t}
    \]
    for some \( C\in\mathbb{R} \).
    By the initial condition, \( x_{i+1}(0)=0 \), so \( C=0 \).
\end{proof}

\begin{lemma}\label{lemma:upper_cascade_bound}
    Under the assumptions of Lemma~\ref{lemma:upper_cascade},
    \begin{equation}
        x_n(t) < pe^{-bt}+p\left(\frac{f}{f+b}\right)^n\left(1-e^{-bt}\right)
    \end{equation}
    for all \( t\in\td \).
\end{lemma}
\begin{proof}
    Assume the hypothesis.
    By equation~\eqref{eq:cascade_ode_3}, the derivative of \( x_n \) is
    \[
        \D{x_n} = fx_{n-1}(t) - bx_n(t).
    \]
    Therefore \( x_n \) has a solution of the form
    \[
        x_n(t) = e^{-bt}\int e^{bt}fx_{n-1}(t)dt.
    \]
    By Lemma~\ref{lemma:upper_cascade}, we know that \( x_{n-1}(t) = \frac{b}{f}F_n(t) \), therefore
    \begin{align*}
        x_n(t)
            &= e^{-bt}\int e^{bt}bF_n(t)dt\\
            &= e^{-bt}\int e^{bt}bp\left(\frac{f}{f+b}\right)^ne^{-(f+b)t}\sum_{k=n}^{\infty}\frac{t^k(f+b)^k}{k!}dt\\
            &= bp\left(\frac{f}{f+b}\right)^ne^{-bt}\int e^{-ft}\sum_{k=n}^{\infty}\frac{t^k(f+b)^k}{k!}dt.
    \end{align*}
    Using the Taylor series of the exponential function, we can rearrange the integral to obtain
    \begin{align*}
        \int e^{-ft}\sum_{k=n}^{\infty}\frac{t^k(f+b)^k}{k!}dt
            &= \int e^{-ft}\left(e^{(f+b)t}-\sum_{k=0}^{n-1}\frac{t^k(f+b)^k}{k!}\right)dt\\
            &= \int e^{bt}dt-\sum_{k=0}^{n-1}\frac{(f+b)^k}{k!}\int t^ke^{-ft}dt.
    \end{align*}
    Since
    \begin{align*}
        \int t^ke^{-ft}dt
            &= -\frac{k!}{f^{k+1}}e^{-ft}\sum_{i=0}^k \frac{t^if^i}{i!}+C_1,
    \end{align*}
    for some \( C_1\in\mathbb{R} \), and
    \begin{align*}
        \int e^{bt}dt
            &= \frac{1}{b}e^{bt}+C_2,
    \end{align*}
    for some \( C_2\in\mathbb{R} \), we see that
    \begin{align*}
        x_n(t)
            &= bp\left(\frac{f}{f+b}\right)^ne^{-bt}\left[\frac{1}{b}e^{bt}-\sum_{k=0}^{n-1}\frac{(f+b)^k}{k!}\left(-\frac{k!}{f^{k+1}}e^{-ft}\sum_{i=0}^k \frac{t^if^i}{i!}\right)+C_3\right]\\
            &= p\left(\frac{f}{f+b}\right)^n+p\frac{b}{f}e^{-bt}\sum_{k=0}^{n-1}\left(\frac{f}{f+b}\right)^{n-k-1}e^{-ft}\sum_{i=0}^k\frac{t^if^i}{i!}+C_4e^{-bt}
    \end{align*}
    for some \( C_3,C_4\in\mathbb{R} \).
    By the initial condition, \( x_n(0)=p \).
    Therefore we can solve for \( C_4 \) in the equation
    \[
        p=p\left(\frac{f}{f+b}\right)^n+p\frac{b}{f}\sum_{k=0}^{n-1}\left(\frac{f}{f+b}\right)^{n-k-1}+C_4,
    \]
    and we see that
    \[
        C_4 = p - p\left(\frac{f}{f+b}\right)^n - p\frac{b}{f}\sum_{k=0}^{n-1}\left(\frac{f}{f+b}\right)^{n-k-1}.
    \]
    After substituting this value for \( C_4 \) into our equation for \( x_n \), we obtain
    \[
    	x_n(t) = pe^{-bt} + p\left(\frac{f}{f+b}\right)^n(1-e^{-bt})+A,
    \]
    where
    \[
    	A = p\frac{b}{f}e^{-bt}\sum_{k=0}^{n-1}\left(\frac{f}{f+b}\right)^{n-k-1}\left(e^{-ft}\sum_{i=0}^k\frac{t^if^i}{i!}-1\right).
    \]
    The lemma immediately follows from the fact that \( A< 0 \).
\end{proof}

At this point, we have derived the solutions and bounds necessary for the cascade of species \( X_0,\ldots,X_n \) from Construction~\ref{const:preprocessor}.
However, we must prove a few lemmas concerning the other two species \( X^\ast \) and \( \Xbar^\ast \) that interact with the top of the cascade.

\begin{construction}\label{const:cascade_top}
    Given \( f>0 \) and \( b>0 \), let \( x,\xbar:\td\rightarrow\td \) be functions that satisfy the ODEs
    \begin{align}
        \D{x} = f\xbar - bx,\label{eq:cascade_top_ode_1}\\
        \D{\xbar} = bx - f\xbar.\label{eq:cascade_top_ode_2}
    \end{align}
\end{construction}

\begin{lemma}\label{lemma:cascade_top}
    If \( x \) and \( \xbar \) are functions constructed according to Construction~\ref{const:cascade_top} with \( f>0 \) and \( b>0 \), then for all \( t\in\td \),
    \begin{align}
        x(t) &= p\left(\frac{f}{f+b}\right)\left(1-e^{-(f+b)t}\right)+x(0)\cdot e^{-(f+b)t}\label{eq:cascade_top_1}\\
        \xbar(t) &= p - x(t),\label{eq:cascade_top_2}
    \end{align}
    where \( p=x(0)+\xbar(0) \).
\end{lemma}
\begin{proof}
    Assume the hypothesis.
    Since \( \D{x}+\D{\xbar}=0 \), it follows that for all \( t\in\td \)
    \[
        x(t)+\xbar(t)=x(0)+\xbar(0),
    \]
        and therefore~\eqref{eq:cascade_top_2} holds.

    To show~\eqref{eq:cascade_top_1} holds, we solve the ODE~\eqref{eq:cascade_top_ode_1} which can be written as
    \[
        \D{x} = f(p-x) - bx,
    \]
    which has solution~\eqref{eq:cascade_top_1}.
\end{proof}

\begin{lemma}\label{lemma:cascade_top_parameters}
    If \( \epsilon\in(0,\frac{1}{2}) \), \( \tau>0 \), and \( x,\xbar \) are constructed according to Construction~\ref{const:cascade_top} with \( p=x(0)+\xbar(0) \), and
    \[
        f \ge \frac{1}{\tau}\log\left(\frac{2p}{\epsilon}\right),\qquad
        b \le f\left(\frac{\epsilon}{2p}\right),
    \]
    then \( x(t) > p-\epsilon \) and \( \xbar(t) < \epsilon \) for all \( t\ge\tau \).
\end{lemma}
\begin{proof}
    Assume the hypothesis.
    Then by Lemma~\ref{lemma:cascade_top}, for all \( t\ge\tau \),
    \begin{align*}
        x(t)
            &\ge p\left(\frac{f}{f+b}\right)\left(1-e^{-(f+b)\tau}\right)+x(0)e^{-(f+b)\tau}\\
            &\ge p\left(\frac{f}{f+b}\right)\left(1-e^{-f\tau}\right).
    \end{align*}
    Since \( f \ge \frac{1}{\tau}\log\left(\frac{2p}{\epsilon}\right) \), \( b \le f\left(\frac{\epsilon}{2p}\right) \), and \( \frac{\epsilon}{2p}<\frac{\epsilon}{2p-\epsilon} \),
    \[
        x(t)
            \ge p\left(\frac{1}{1+\frac{\epsilon}{2p}}\right)\left(1-\frac{\epsilon}{2p}\right)
            = p\left(1-\frac{\epsilon}{2p}\right)^2
            > p-\epsilon.\qedhere
    \]
\end{proof}

\subsection{Proof of Input Enhancement Theorem}\label{sub:cleaner_proof}
We now have the machinery that we need to prove the Input Enhancement Theorem.

\begin{proof}[Proof of Theorem~\ref{theorem:preprocessor}]
    Assume the hypothesis.
    Then \( \tau>0 \), \( \epsilon\in(0,\frac{1}{2}) \), \( \bdelta=(\delta_u,\delta_h,\delta_0,\delta_k) \) with \( \delta_u\in(0,\frac{1}{3}) \), \( \delta_h\in(0,\epsilon) \), \( \delta_0\in(0,\frac{1}{2}) \), \( \delta_k>0 \), and \( N\supX=N^\supX(\tau,\epsilon,\bdelta) \) and \( \xn^\supX \) are constructed according to Construction~\ref{const:preprocessor}.
    We now must show that \( N^\supX,\xn^\supX\models_\epsilon^{\bdelta}\Phi^\supX(\tau) \).

    Now let \( n=|S|-2 \),
    let \( \bc=(\bu,V,h) \) be a context satisfying \( \alpha(\bc) \),
    let \( \bchat=(\buhat,V,\hhat) \) be \( (\delta_u,\delta_h) \)-close to \( \bc \),
    let \( \xnhat \) be \( \delta_0 \)-close to \( \xn^\supX \),
    let \( \Nhat \) be \( \delta_k \)-close to \( N^\supX \), and
    let \( p \) and \( p^\ast \) be the constants
    \[
        p = \sum_{i=0}^n \xnhat(X_i),\hspace*{4em}
        p^\ast = \xnhat(X^\ast)+\xnhat(\Xbar^\ast).
    \]

    It now suffices to show that \( \Nhat_{\bchat,\xnhat} \) is \( \epsilon \)-close to a function \( \bv\in\mathcal{C}[V] \) that satisfies \( \phi(\bu,\bv) \), \ie, if \( (b,I) \) is an input event for \( \bu \), then \( (b,I_\tau) \) must be an output event for \( \bv \).
    We prove this in two cases corresponding to \( b=1 \) and \( b=0 \) by invoking many of the lemmas from the previous section.

    The state species of \( \Nhat \) are naturally split up into two parts.
    The first part is the cascade of species \( X_0,\ldots,X_n \), and the second part are the species \( X^\ast,\Xbar^\ast \) which are affected by the top of the cascade.
    The ODEs for species \( X_0,\ldots,X_n \) of \( \Nhat \) can be derived from the reactions in Construction~\ref{const:preprocessor} along with the perturbed mass action function from equation~\eqref{eq:perturbated_ode} and are
    \begin{align}
        \D{x_0} &= \sum_{i=1}^n\hat{k}_1x_i-(\hat{k}_1x)x_0,\\
        \D{x_i} &= (\hat{k}_1x)x_{i-1}-(\hat{k}_1x+\hat{k}_1)x_i\quad\text{for }0<i<n,\\
        \D{x_n} &= (\hat{k}_1x)x_{n-1}-\hat{k}_1x_n.
    \end{align}
    Similarly, the ODEs for \( X^\ast \) and \( \Xbar^\ast \) are
    \begin{align}
        \D{x^\ast} &= (\hat{k}_2x_n)\xbar^\ast-\hat{k}_2x^\ast,\label{eq:x_ast_ode}\\
        \D{\xbar^\ast} &= \hat{k}_2x^\ast - (\hat{k}_2x_n)\xbar^\ast.\label{eq:xbar_ast_ode}
    \end{align}

    Since \( \D{x^\ast}+\D{\xbar^\ast}=0 \), it is easy to show that \( x^\ast(t)+\xbar^\ast(t)=p^\ast \) for all \( t\in\td \).
    Similarly, \( \sum_{i=0}^n x_i(t)=p \) for all \( t\in\td \).

    Let \( (1,I) \) be an input event for \( \bu \), where \( I=[t_1,t_2] \).
    Since the input signal can be perturbed by \( \delta_u \), it follows that \( x(t)>1-\delta_u \) for all \( t\in I \).
    We also know that the rate constants can be perturbed by \( \delta_k \).
    To minimize the concentration of \( X_n \) in the interval \( I \), we assume that all the concentration of \( X_0,\ldots,X_n \) is in \( X_0 \) at time \( t_1 \).
    We also maximize the rate of falling down the cascade and minimize the rate of climbing the cascade.

    Therefore by Lemma~\ref{lemma:lower_cascade}, for all \( t\in I \),
    \[
        x_n(t)>p\left(\frac{f}{f+b}\right)^n\sum_{i=n}^{\infty}\frac{t^i(f+b)^i}{i!}e^{-(f+b)(t-t_1)},
    \]
    where \( f=(k_1-\delta_k)(1-\delta_u) \) and \( b=k_1+\delta_k \).
    Since \( x_n \) is monotonically increasing, for all \( t\in[t_1+\frac{\tau}{2},t_2] \), \( x_n(t)\ge x_n(\frac{\tau}{2}) \), and therefore
    \[
        x_n(t)>p\left(\frac{f}{f+b}\right)^n\sum_{i=n}^{\infty}\frac{t^i(f+b)^i}{i!}e^{-(f+b)\frac{\tau}{2}}.
    \]
    By Lemma~\ref{lemma:lower_cascade_bound}, for all \( t\in[t_1+\frac{\tau}{2},t_2] \),
    \[
        x_n(t)>p\left(\frac{f}{f+b}\right)^n\left(1-ne^{-\frac{1}{n}(f+b)\frac{\tau}{2}}\right).
    \]
    Since \( k_1>\delta_k+\frac{2n}{\tau(1-\delta_u)}\log(2n) \), Corollary~\ref{corollary:lower_cascade_bound} tells us
    \begin{align*}
        x_n(t)
            &>p\left(\frac{f}{f+b}\right)^n\left(\frac{1}{2}\right)\\
            &=\frac{p}{2}\left(\frac{(k_1-\delta_k)(1-\delta_u)}{(k_1-\delta_k)(1-\delta_u)+k_1+\delta_k}\right)^n\\
            &=\frac{p}{2}\left(\frac{1-\delta_u}{1-\delta_u+u}\right)^n,
    \end{align*}
    where \( u=\frac{k_1+\delta_k}{k_1-\delta_k} \).
    Since \( k_1>2\frac{\delta_k(2+\delta_u)}{\delta_u} \), we know that \( u < 1+\delta_u \) and therefore
    \[
        x_n(t) > \frac{p}{2}\left(\frac{1-\delta_u}{2}\right)^n.
    \]
    Since the initial condition can be perturbed by at most \( \delta_0 \), \( p>\frac{10}{\epsilon-\delta_h}\left(\frac{2}{1-\delta_u}\right)^n \), therefore
    \[
        x_n(t) > \frac{5}{\epsilon-\delta_h},
    \]
    for all \( t\in[t_1+\frac{\tau}{2},t_2] \).

    Recall that the ODEs for \( X^\ast \) and \( \Xbar^\ast \) are~\eqref{eq:x_ast_ode} and~\eqref{eq:xbar_ast_ode}.
    To minimize the concentration of \( X^\ast \) in the interval \( [t_1+\frac{\tau}{2},t_2] \), we minimize the production of \( X^\ast \) and maximize the production of \( \Xbar^\ast \).
    By Lemma~\ref{lemma:cascade_top}, for all \( t\in[t_1+\tau,t_2] \),
    \[
        x^\ast(t) > p^\ast\left(\frac{f}{f+b}\right)\left(1-e^{-(f+b)\frac{\tau}{2}}\right),
    \]
    where \( f=(k_2-\delta_k)\frac{5}{\epsilon-\delta_h} \) and \( b=k_2+\delta_k \).
    Since \( k_2>4\delta_k \), then \( \frac{k_2+\delta_k}{k_2-\delta_k}<\frac{5}{3} \) and therefore
    \[
        b=k_2+\delta_k<\frac{\epsilon-\delta_h}{3}\left[(k_2-\delta_k)\frac{5}{\epsilon-\delta_h}\right]=\frac{\epsilon-\delta_h}{3}f,
    \]
    whence
    \[
        b<\frac{\epsilon-\delta_h}{2p^\ast}f.
    \]
    Since \( \frac{5}{\epsilon-\delta_h}>1 \) and \( k_2=\frac{2}{\tau}\log\left(\frac{3}{\epsilon-\delta_h}\right)+4\delta_k \),
    \[
        f
            = (k_2-\delta_k)\frac{5}{\epsilon-\delta_h}
            > \frac{2}{\tau}\log\left(\frac{3}{\epsilon-\delta_h}\right)
            > \frac{2}{\tau}\log\left(\frac{2p^\ast}{\epsilon-\delta_h}\right).
    \]
    By Lemma~\ref{lemma:cascade_top_parameters}, for all \( t\in[t_1+\tau,t_2] \),
    \[
        x^\ast(t)>p^\ast-\epsilon+\delta_h.
    \]
    Since the initial state can only be perturbed by at most \( \delta_0 \) and the output function can only introduce \( \delta_h \) error.
    It immediately follows that
    \[
        N_{\bchat,\xnhat}(t)>1-\epsilon.
    \]
    Therefore \( \Nhat_{\bchat,\xnhat}(t) \) is \( \epsilon \)-close to satisfying the requirement that \( (1,I_\tau) \) is an output event.

    It remains to be shown that \( \Nhat_{\bchat,\xnhat} \) is \( \epsilon \)-close to handling input events of the form \( (0,I) \).
    To show this, let \( (0,I) \) be an input event, and let \( I=[t_1,t_2] \).
    Therefore \( x(t)<\delta_u \) for all \( t\in I \).
    Similar to the above argument, by Lemma~\ref{lemma:upper_cascade_bound}, for all \( t\in I \),
    \[
        x_n(t) < pe^{-b(t-t_1)}+p\left(\frac{f}{f+b}\right)^n\left(1-e^{-b(t-t_1)}\right)
    \]
    where \( f=(k_1+\delta_k)\delta_u \) and \( b=k_1-\delta_k \).
    Since this function is monotonically decreasing, for all \( t\in[t_1+\frac{\tau}{2},t_2] \),
    \begin{align*}
        x_n(t)
            &< p\left(\frac{f}{f+b}\right)^n+pe^{-b\frac{\tau}{2}}\\
            &= p\left(\frac{(k_1+\delta_k)\delta_u}{(k_1+\delta_k)\delta_u+k_1-\delta_k}\right)^n
                +pe^{-b\frac{\tau}{2}}\\
            &= p\left(\frac{\delta_u}{\delta_u+u}\right)^n
                +pe^{-b\frac{\tau}{2}},
    \end{align*}
    where \( u=\frac{k_1-\delta_k}{k_1+\delta_k} \).
    Since \( k_1>\frac{\delta_k(2-\delta_u)}{\delta_u} \), we know that \( u>1-\delta_u \), whence for all \( t\in[t_1+\frac{\tau}{2},t_2] \),
    \[
        x_n(t)< p\delta_u^n+pe^{-b\frac{\tau}{2}}.
    \]
    Since \( p<\frac{10}{\epsilon-\delta_h}\left(\frac{2}{1-\delta_u}\right)^n+2\delta_0 \),
    \begin{align*}
        x_n(t)
            &< \frac{10}{\epsilon - \delta_h}\left(\frac{2\delta_u}{1-\delta_u}\right)^n+\delta_u^n+pe^{-b\frac{\tau}{2}}\\
            &< \frac{10+\epsilon - \delta_h}{\epsilon - \delta_h}\left(\frac{2\delta_u}{1-\delta_u}\right)^n+pe^{-b\frac{\tau}{2}}\\
            &< \frac{32}{3(\epsilon - \delta_h)}\left(\frac{1-\delta_u}{2\delta_u}\right)^{-n}+pe^{-b\frac{\tau}{2}}.
    \end{align*}
    Since \( n\ge\log_{\left(\frac{1-\delta_u}{2\delta_u}\right)}\left(\frac{64}{(\epsilon-\delta_h)^2}\right) \),
    \[
        x_n(t)
            < \frac{32}{3(\epsilon-\delta_h)}\left(\frac{(\epsilon-\delta_h)^2}{64}\right)+pe^{-b\frac{\tau}{2}}
            = \frac{\epsilon-\delta_h}{6}+pe^{-b\frac{\tau}{2}}.
    \]
    As we showed before, \( p<\frac{32}{3(\epsilon-\delta_h)}\left(\frac{2}{1-\delta_u}\right)^n+2\delta_0 \), whence
    \[
        x_n(t)<\frac{\epsilon-\delta_h}{6}+\frac{32}{3(\epsilon-\delta_h)}\left(\frac{4}{1-\delta_u}\right)^ne^{-b\frac{\tau}{2}} + 2\delta_0e^{-b\frac{\tau}{2}}.
    \]
    Since \( b=k_1-\delta_k>\frac{2}{\tau}\log\left(\frac{640}{(\epsilon-\delta_h)^2}\left(\frac{2}{1-\delta_u}\right)^n\right) \),
    \begin{align*}
        x_k(t)
            &<\frac{\epsilon-\delta_h}{6}+\frac{\epsilon-\delta_h}{60}+2\delta_0e^{-b\frac{\tau}{2}}
            <\frac{\epsilon-\delta_h}{6}+\frac{\epsilon-\delta_h}{30},
    \end{align*}
    whence for all \( t\in[t_1+\frac{\tau}{2},t_2] \)
    \[
        x_n(t)<\frac{\epsilon-\delta_h}{5}.
    \]

    We now bound the concentration of \( X^\ast \) and \( \Xbar^\ast \).
    If \( f=(k_2+\delta_k)\frac{\epsilon-\delta_h}{5} \) and \( b=k_2-\delta_k \), then by Lemma~\ref{lemma:cascade_top}
    \[
        \xbar^\ast(t)>p\left(\frac{f}{f+b}\right)\left(1-e^{-(f+b)\frac{\tau}{2}}\right),
    \]
    for all \( t\in[t_1+\tau,t_2] \).
    Since \( \frac{k_2-\delta_k}{k_2+\delta_k}>\frac{3}{5} \),
    \[
        f=(k_2+\delta_k)\frac{\epsilon-\delta_h}{5}<\frac{\epsilon-\delta_h}{3}(k_2-\delta_k)<\frac{\epsilon-\delta_h}{2p^\ast}b.
    \]
    Since \( b=k_2-\delta_k>\frac{2}{\tau}\log\left(\frac{3}{\epsilon-\delta_h}\right)>\frac{2}{\tau}\log\left(\frac{2p^\ast}{\epsilon-\delta_h}\right) \), by Lemma~\ref{lemma:cascade_top_parameters}, for all \( t\in[t_1+\tau,t_2] \),
    \[
        \xbar^\ast(t)>p^\ast-\epsilon+\delta_h.
    \]
    Since \( x^\ast(t)+\xbar^\ast(t)=p^\ast \) for all \( t\in\td \), it follows that \( x^\ast(t)<\epsilon-\delta_h \) for all \( t\in I_\tau \).
    Since \( p^\ast>1+\delta_0 \), it follows that \( \xbar^\ast(t)>1-\epsilon+\delta_h \) for all \( t\in I_\tau \).

    Finally, since the output function can at most deviate by \( \delta_h \) from the solutions of \( x^\ast(t) \) and \( \xbar^\ast(t) \), it is clear that \( \Nhat_{\bchat,\xnhat} \) is \( \epsilon \)-close to having \( (0,I_\tau) \) as a valid output event.
\end{proof}

    \section{Proof of Initialization Lemma}\label{app:initialization}
\begin{proof}[Proof of Lemma~\ref{lemma:initialization}]
    In this proof, we must show that the I/O CRN \( N \), when initialized with its initial state \( \xn \), will still be \( \delta_0 \)-encoding its set of initial states \( I \) at time \( t=\frac{1}{2} \).
    Before we begin the argument, we must fix the arbitrary perturbations of the I/O CRN.
    Since this lemma applies to both the base case and the induction step of our overall argument, we fix an arbitrary input string \( w\in\Sigma^\ast \).
    Now let \( \bc=(\bu,V,h) \) be a context satisfying \( \alpha_w(\bc) \),
    let \( \bchat=(\buhat,V,h) \) be \( (\delta_u,0) \)-close to \( \bc \),
    let \( \xnhat \) be \( \delta_0 \)-close to \( \xn \),
    let \( \Nhat \) be \( \delta_k \)-close to \( N \),
    and let \( \bxhat(t) \) be the unique solution of \( \Nhat \) when initialized to \( \xnhat \).
    To complete the proof we must show that \( Y_Q \) \( \eta \)-encodes \( I \) in the state \( \bxhat(\frac{1}{2}) \).
    In other words, for each \( q\in I \) we must show that \( \yq(\frac{1}{2}) > \pyq - \eta \) and for each \( q\not\in I \) that \( \yq(\frac{1}{2})<\eta \).

    For the first part, let \( q\in I \).
    Recall that the ODE for \( \Yq \) is
    \begin{align*}
        \D{\yq}
            &= \khat_2\xc\zq\yqbar - \khat_2\xc\zqbar\yq + \khat_2\yq^2\yqbar - \khat_2\yq\yqbar^2\\
            &> \khat_2\yq^2\yqbar - \khat_2\yq\yqbar^2 - \khat_2\delta_0\zqbar\yq.
    \end{align*}
    Since \( \khat_2 \) is \( \delta_k \)-close to the constant \( k_2 \), we have
    \begin{align*}
        \D{\yq}
            &> (k_2-\delta_k)\yq^2\yqbar - (k_2+\delta_k)\yq\yqbar^2 - (k_2+\delta_k)\delta_0\zqbar\yq.
    \end{align*}
    Since the sum of the concentrations of \( \Zq \) and \( \Zqbar \) is the constant \( \pzq \), we know that \( \zqbar(t) \) is bounded by \( \pzq \).
    Thus,
    \begin{align*}
        \D{\yq}
            &> (k_2-\delta_k)\yq^2\yqbar - (k_2+\delta_k)\yq\yqbar^2 - (k_2-\delta_k)\delta_0\pzq\yq.
    \end{align*}
    Since \( \yqbar=\pyq-\yq \), we can simplify the ODE to
    \begin{align*}
        \D{\yq} >
            &(k_2-\delta_k)\yq^2(\pyq-\yq) - (k_2+\delta_k)\yq(\pyq-\yq)^2\\
            &- (k_2-\delta_k)\delta_0\pzq\yq.
    \end{align*}
    At this point, everything in the ODE is a constant except for the function \( \yq \).
    If we let \( a \), \( b \), \( c \), and \( p \) be the constants
    \begin{align*}
        a&=k_2-\delta_k,\qquad
        b=k_2+\delta_k,\qquad
        c=(k_2-\delta_k)\pzq\delta_0,\qquad
        p=\pyq,
    \end{align*}
    then we can rewrite the ODE in a simpler form
    \[
        \D{\yq} > a\yq^2(p-\yq) - b\yq(p-\yq)^2 - c\yq.
    \]
    The above ODE has identical structure to that of the termolecular signal restoration algorithm from~\cite{cKlin16}.
    This means that if the inequality \( c < \frac{p^2a^2}{4(a+b)} \) holds, we can make use of Theorem~3.2 from~\cite{cKlin16} to bound the concentration of \( \Yq \) during the interval \( [0,\frac{1}{2}] \).
    It is routine to verify this and can easily be shown using the facts that \( k_2>25 \), \( 1-\delta_0 < \pzq < 1+\delta_0 \), and \( \delta_0,\delta_k < \frac{1}{20} \).

    At this point, we know that during the interval \( [0,\frac{1}{2}] \) the behavior of \( \Yq \) is bounded by the termolecular signal restoration algorithm and that the constant \( c \) is small enough to introduce bistability to the system.
    We now show that the concentration of \( \Yq \) is attracted to the stable fixed point close to 1 and therefore remains unaffected.

    Let \( E_1 \) and \( E_2 \) be the constants
    \begin{align}
        E_1&=p\left(\frac{b}{a+b}\right)+A,\\
        E_2&= p - A,
    \end{align}
    where \( A=\frac{p}{2}\left(\frac{a}{a+b}\right)\left(1-\sqrt{1-c^\ast}\right) \) and \( c^\ast=c\cdot\frac{4(a+b)}{p^2a^2} \).
    These constants are two of the equilibrium points of the signal restoration algorithm mentioned previously.
    Because of the stability of these points shown in~\cite{cKlin16}, if \( \yq(0)>E_1 \), then \( \Yq \) will converge to \( E_2 \), whereas if \( \yq(0)<E_1 \) then it would converge to 0.
    It is routine to verify that \( \yq(0) > E_1 \) but can be easily shown using the bounds mentioned previously along with \( \yq(0)>1-\delta_0 \).

    Since \( E_1 \) is the decision point of the signal restoration algorithm, Theorem~3.2 from~\cite{cKlin16} tells us that the concentration of \( \Yq \) will converge away from the constant \( E_1 \) to the constant \( E_2 \).
    It immediately follows from the fact that \( E_1 < \yq(0) < E_2 \) that \( \yq(t) \ge \yq(0) > 1-\delta_0 \) for all \( t\in[0,\frac{1}{2}] \).

    It remains to be shown that if \( q\not\in I \), then \( \yq(\frac{1}{2}) < \delta_0 \).
    This immediately follows by the symmetry imposed by the dual relationship of \( Y_q \) and \( \Yqbar \).
\end{proof}
    \section{Proof of State Restoration Lemma}\label{app:state_restoration}
\begin{proof}[Proof of Lemma~\ref{lemma:state_restoration}]
    In this proof, we must show that the I/O CRN \( N \), when initialized with \( \xn \), is capable of maintaining and improving its encoding of a set \( A\subseteq Q \).
    Before we begin the argument, we must fix the arbitrary perturbations of the I/O CRN.
    Since this lemma also applies to both the base case and the induction step of our overall argument, we fix an arbitrary input string \( w\in\Sigma^\ast \).
    Now let \( \bc=(\bu,V,h) \) be a context satisfying \( \alpha_w(\bc) \),
    let \( \bchat=(\buhat,V,h) \) be \( (\delta_u,0) \)-close to \( \bc \),
    let \( \xnhat \) be \( \delta_0 \)-close to \( \xn \),
    let \( \Nhat \) be \( \delta_k \)-close to \( N \),
    and let \( \bxhat(t) \) be the unique solution of \( \Nhat \) when initialized to \( \xnhat \).

    We also assume the hypothesis of Lemma~\ref{lemma:state_restoration}, \ie, that \( Y_Q \) \( \frac{1}{20} \)-encodes the set \( A\subseteq Q \) in state \( \bxhat(t_1) \) and \( x^\ast(t)\le\gamma \) for all \( t\in[t_1,t_2] \) and for each \( X^\ast\in U^\ast \).
    To complete the proof, we must show that \( Y_Q \) \( \eta \)-encodes \( A \) for all \( t\in [t_1+\frac{1}{2}, t_2] \).

    Let \( q\in A \).
    Recall that the ODE for \( \Yq \) is
    \[
        \D{\yq} = \khat_2\xc\zq\yqbar - \khat_2\xc\zqbar\yq + \khat_2\yq^2\yqbar - \khat_2\yq\yqbar^2.
    \]
    We will be examining the behavior of \( Y_q \) during the interval \( [t_1, t_2] \).
    During this interval, we know by the hypothesis that \( x^\ast(t)\le\gamma \) for each \( X^\ast\in U^\ast \).
    Thus we know that during the interval \( [t_1,t_2] \)
    \begin{align*}
        \D{\yq}
            &> \khat_2\yq^2\yqbar - \khat_2\yq\yqbar^2 - \khat_2\gamma\zqbar\yq,\\
            &> (k_2-\delta_k)\yq^2(\pyq-\yq) - (k_2+\delta_k)\yq(\pyq-\yq)^2\\
            &\quad- (k_2+\delta_k)\gamma\pzq\yq.
    \end{align*}
    Now if we define the constants \( a \), \( b \), \( c \), and \( p \) to be
    \begin{align*}
        a&=k_2-\delta_k,\qquad
        b=k_2+\delta_k,\qquad
        c=(k_2+\delta_k)\pzq\gamma,\qquad
        p=\pyq,
    \end{align*}
    then we can rewrite the above ODE as
    \[
        \D{\yq} > a\yq^2(p-\yq) - b\yq(p-\yq)^2 - c\yq.
    \]
    We also define the constants \( E_1 \) and \( E_2 \)
    \begin{align}
        E_1&=p\left(\frac{b}{a+b}\right)+A,\\
        E_2&= p - A,
    \end{align}
    where \( A=\frac{p}{2}\left(\frac{a}{a+b}\right)\left(1-\sqrt{1-c^\ast}\right) \) and \( c^\ast=c\cdot\frac{4(a+b)}{p^2a^2} \).

    It is easy to verify that \( c<\frac{p^2a^2}{4(a+b)} \) and \( E_1<\yq(t_1) \), so Theorem~3.2 from~\cite{cKlin16} tells us that \( \yq(t) \) is converging toward the value \( E_2 \).

    We now show that the constant \( E_2 \) is sufficiently high to restore the concentration of \( \Yq \) to at least \( p(\Yq)-\eta \).
    Using the definition of \( \gamma \) in~\eqref{eq:gamma_specification}, \( k_2 \) in~\eqref{eq:k2_specification}, and the fact \( \delta_k,\delta_0<\frac{1}{20} \), it is not difficult to show that:
    \begin{equation}\label{eq:greater_than_eta_8}
        A 
            < \frac{2c}{pa} 
            = \frac{2\gamma(k_2+\delta_k)\pzq}{\pyq(k_2-\delta_k)}
            < 2\gamma\left(\frac{1+\frac{\delta_k}{k_2}}{1-\frac{\delta_k}{k_2}}\right)\left(\frac{1+\delta_0}{1-\delta_0}\right)
            < \frac{\eta}{8}.
    \end{equation}
    It follows that
    \begin{equation}
        E_2 = \pyq - A > \pyq-\frac{\eta}{8}.
    \end{equation}

    By Theorem~3.3 in~\cite{cKlin16}, the amount of time \( \Delta t \) it takes the signal restoration reactions to restore the concentration of \( \Yq \) to \( \pyq-\eta \) from \( \pyq-\delta_0 \) is bounded by
    \begin{equation}\label{eq:restore_time_bound}
        \Delta t < \frac{a+b}{abp^2(1-c\frac{4(a+b)}{p^2a^2})}\log u,
    \end{equation}
    where
    \[
       u=\frac{(p-\eta-E_1)(E_2-p+\delta_0)}{(p-\delta_0-E_1)(E_2-p+\eta)}.
    \]
    Using previous definitions and bounds, it is routine to verify that the right-hand side of~\eqref{eq:restore_time_bound} is bounded by \( \frac{1}{2} \).

    It follows that within \( \frac{1}{2} \) time, the concentration of \( \Yq \) reaches \( \pyq-\eta \), and therefore for all \( t\in[t_1+\frac{1}{2},t_2] \), \( \yq(t)>\pyq-\eta \).

    This finishes one half of the proof, namely, that if \( q\in A \) that the I/O CRN \( N \) robustly keeps the value of \( \Yq \) \( \eta \)-close to \( \pyq \) during the interval \( [t_1+\frac{1}{2},t_2] \).
    It remains to be shown that for \( q\not\in A \), \( \yq(t)<\eta \) for all \( t\in[t_1+\frac{1}{2},t_2] \).
    This follows by the symmetry of \( \Yq \) and its dual \( \Yqbar \).
\end{proof}

    \section{Proofs of Computation Lemmas}\label{app:computation_lemmas}
In this appendix, we prove Lemmas~\ref{lemma:reset},~\ref{lemma:transition},~\ref{lemma:copyback}.
Recall that there are variables defined in \cref{sec:nfa_sim} that are relevant to these lemmas, and we repeat them here for readability.
Let \( w\in\Sigma^\ast \) and \( a\in\Sigma \) and assume the inductive hypothesis \( N,\xn\models_\eta^{\bdelta^\ast}\Phi_w \) of Lemma~\ref{lemma:induction_step} holds.
Let \( \bc=(\bu,V,h) \) be a context satisfying \( \alpha_{wa}(\bc) \),
let \( \bchat=(\buhat,V,h) \) be \( (\delta_u,0) \)-close to \( \bc \),
let \( \xnhat \) be \( \delta_0 \)-close to \( \xn \), and
let \( \Nhat \) be \( \delta_k \)-close to \( N \).

Let \( I=[b,b+12] \) be the final symbol event of the input \( \bu \).
Then \( I \) is an \( a \)-event and \( \tau(\bu)=b+12 \).
The proofs of computation lemmas involves closely examining the behavior of the I/O CRN \( N \) during the three pulses of this \( a \)-event s shown in Figure~\ref{fig:a_event_timing_graph_appendix}:

\begin{figure}
    \centering
    \includegraphics{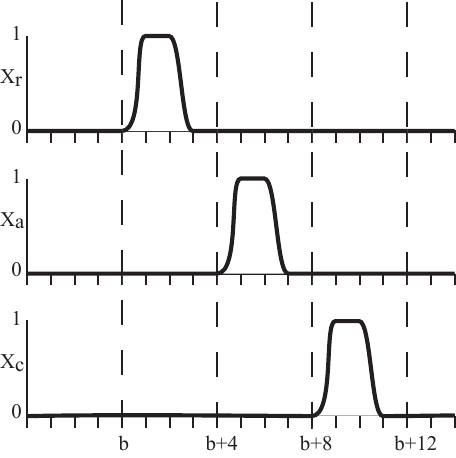}
    \caption{\label{fig:a_event_timing_graph_appendix}
        The \( X_r \)-, \( X_a \),- and \( X_c \)-pulses of the final \( a \)-event.
    }
\end{figure}

Before we start proving the computation lemmas, we first state and prove some helpful observations.

\begin{observation}\label{obs:yq_encodes_first_pulse}
    \( Y_Q \) \( \eta \)-encodes \( \deltaof{w} \) during the interval \( [b,b+8] \).
\end{observation}
\begin{proof}
    Let \( \bu^\ast \) be a terminal input such that \( \bu^\ast(t)=\bu(t) \) for all \( t\in[0,b] \) and \( \bu^\ast(t)=\bm{0} \) for all \( t>b \).
    Then \( w(\bu^\ast)=w \) and \( \tau(\bu^\ast)\le b \).
    Since \( N,\xn\models_\eta^{\bdelta^\ast}\Phi_w \) by our inductive hypothesis, we know that \( Y_Q \) \( \eta \)-encodes \( \deltaof{w} \) starting at time \( \tau(\bu^\ast) \).
    Since \( \bu \) and \( \bu^\ast \) agree at every \( t\in[0,b] \), the concentrations of \( \Yq \) for each \( q\in Q \) must also agree at every time \( t\in[0,b] \).
    Therefore \( Y_Q \) \( \eta \)-encodes \( \deltaof{w} \) at time \( b \).
    Finally, the input species \( X_c \) is below \( \delta_u \) during the interval \( [b,b+8] \), so the values of the species in \( Y_Q \) will be maintained by the state restoration reactions as shown in the proof of Lemma~\ref{lemma:state_restoration} in Appendix~\ref{app:state_restoration}.
\end{proof}

\begin{observation}
    \( x_r^\ast(b),x_c^\ast(b) < \gamma \).
\end{observation}
\begin{proof}
    This follows from the fact that the concentrations of inputs \( X_r \) and \( X_c \) are less than \( \delta_u \) in the interval \( [b-\frac{1}{2}, b] \).
    Therefore the preprocessed input species \( X_r^\ast \) and \( X_c^\ast \) have time to drop below \( \gamma \) before time \( b \).
\end{proof}

\subsection{Proof of Reset Lemma}
\begin{proof}[Proof of Lemma~\ref{lemma:reset}]
    For \( q\in Q \), the ODE for \( \Zq \) is
    \begin{align*}
        \D{\zq}
            &= -\khat_1\xr\zq + \sum_{(s,a,q)\in\Delta}\khat_1\xa\ys\zqbar\\
            &< -(k_1-\delta_k)\xr\zq + \sum_{(s,a,q)\in\Delta}(k_1+\delta_k)\xa p(Y_r)\zqbar.
    \end{align*}
    The interval \( [b,b+4] \) is an \( X_r \)-pulse and therefore every species \( X_a^\ast \) for \( a\in\Sigma \) must have concentration \( \gamma \)-close to zero.
    Therefore
    \begin{align*}
        \D{\zq}
            &< -(k_1-\delta_k)\xr\zq + \sum_{(s,a,q)\in\Delta}p(Y_r)(k_1+\delta_k)\gamma\zqbar\\
            &< -(k_1-\delta_k)\xr\zq + |Q|(1+\delta_0)(k_1+\delta_k)\gamma\zqbar.
    \end{align*}
    During the interval \( [b+1,b+2] \), the species \( X_r \) is \( \delta_u \)-close to one.
    Therefore during the interval \( [b+1.5, b+2] \) the species \( X^\ast_r \) is \( \gamma \)-close to 1.
    Thus
    \begin{align*}
        \D{\zq}
            &< -(k_1-\delta_k)(1-\gamma)\zq + |Q|(1+\delta_0)(k_1+\delta_k)\gamma\zqbar.
    \end{align*}
    Let \( \hat{f} \), \( \hat{b} \), and \( \hat{p} \) be constants defined by
    \begin{align*}
        \hat{f} &= |Q|(1+\delta_0)(k_1+\delta_k)\gamma\hspace*{3em}
        \hat{b} = (k_1-\delta_k)(1-\gamma),\hspace*{3em}
        \hat{p} = p(\Zq),
    \end{align*}
    then we can rewrite the above ODE as
    \[
        \D{\zq} < \hat{f}\zqbar - \hat{b}\zq,
    \]
    which has identical structure to the ODE from Construction~\ref{const:cascade_top}.
    By Lemma~\ref{lemma:cascade_top}, we have the bound
    \begin{align*}
        \zq(b+2)
            &< \hat{p}\left(\frac{\hat{f}}{\hat{f}+\hat{b}}\right)\left(1-e^{-(\hat{f}+\hat{b})\frac{1}{2}}\right)+\zq(b+1.5)\cdot e^{-(\hat{f}+\hat{b})\frac{1}{2}}\\
            &< \hat{p}\left(\frac{\hat{f}}{\hat{f}+\hat{b}}\right)\left(1-e^{-(\hat{f}+\hat{b})\frac{1}{2}}\right)+\hat{p}e^{-(\hat{f}+\hat{b})\frac{1}{2}}\\
            &= \hat{p}\left(\frac{\hat{f}}{\hat{f}+\hat{b}}\right)+\hat{p}\left(\frac{\hat{b}}{\hat{f}+\hat{b}}\right)e^{-(\hat{f}+\hat{b})\frac{1}{2}}
            < \hat{p}\hat{f}+\hat{p}e^{-\frac{\hat{b}}{2}}.
    \end{align*}
    Using the definition of \( k_1 \) from equation~\eqref{eq:k1_specification}, it is easy to show that \( k_1>\delta_k+\frac{2}{1-\gamma}\log\left(\frac{4(1+\delta_0)}{\eta}\right) \), therefore
    \begin{align*}
        \zq(b+2)
            &< \hat{p}\hat{f}+\frac{\eta}{4}.
    \end{align*}
    It is routine but easy to verify that \( \hat{p}\hat{f}>\frac{\eta}{4} \), and so \( \zq(b+2) < \frac{\eta}{2} \).
    
    During the interval \( [b+2,b+4] \), the derivative of \( \Zq \) is bounded by
    \begin{align*}
        \D{\zq}
            &< \hat{f}(\hat{p}-\zq),
            < \hat{p}\hat{f},
            < \frac{\eta}{4},
    \end{align*}
    which means less than \( \frac{\eta}{2} \) of \( \Zq \) is produced over the interval \( [b+2,b+4] \).
    Therefore \( \zq(b+4)<\eta \).
    Since \( q \) was arbitrary, this means that \( Z_Q \) \( \eta \)-encodes the set \( \emptyset \) at time \( b+4 \).
\end{proof}

\subsection{Proof of Transition Lemma}
\begin{proof}[Proof of Lemma~\ref{lemma:transition}]
    We prove this in two steps.
    First we prove that if \( q\in\deltaof{wa} \), then for all \( t\in[b+8,b+12] \) \( \zq(t)> p(\Zq)-\frac{1}{20} \), and second we prove that if \( q\not\in\deltaof{wa} \), then for all \( t\in[b+8,b+12] \) \( \zq(t)<\frac{1}{20} \).
    We also depend on the fact that Lemma~\ref{lemma:reset} states that all the portal species have been reset to a concentration less than \( \eta \) at time \( b+4 \).

    For the first part, let \( q\in\deltaof{wa} \).
    Then there exists a state \( s\in\deltaof{w} \) such that \( (s,a,q)\in\Delta \).
    This means that there is at least one reaction from equation~\eqref{eq:transition_reaction} and Construction~\ref{const:crn_translation} that computes the transition \( (s,a,q)\in\Delta \).
    Therefore we can bound the ODE corresponding to \( Z_q \) by
    \begin{align*}
       \D{\zq}
            &= -\khat_1\xr\zq + \sum_{(s,a,q)\in\Delta}\khat_1\xa\ys\zqbar,\\
            &> -\khat_1\xr\zq + \khat_1\xa\ys\zqbar,\\
            &> -(k_1+\delta_k)\xr\zq + (k_1-\delta_k)\xa\ys\zqbar.
    \end{align*}
    During the interval \( [b+5,b+6] \), the input signal is at the peak of the \( X_a \)-pulse, and therefore during the interval \( [b+5.5,b+6] \) we know that
    \[
        \D{\zq} > -(k_1+\delta_k)\gamma\zq + (k_1-\delta_k)(1-\gamma)\ys\zqbar.
    \]
    By Observation~\ref{obs:yq_encodes_first_pulse}, the set \( Y_Q \) \( \eta \)-encodes \( \deltaof{w} \) during \( [b,b+8] \), and since \( s\in\deltaof{w} \) we know that
    \begin{align*}
        \D{\zq}
            &> -(k_1+\delta_k)\gamma\zq + (k_1-\delta_k)(1-\gamma)(1-\eta)\zqbar.
    \end{align*}
    Now let \( \hat{f} \), \( \hat{b} \), and \( \hat{p} \) be the constants
    \begin{align*}
        \hat{f} = (k_1-\delta_k)(1-\gamma)(1-\eta)\hspace*{3em}
        \hat{b} = (k_1+\delta_k)\gamma\hspace*{3em}
        \hat{p} = p(\Zq),
    \end{align*}
    so that
    \[
        \D{\zq} > \hat{f}\zqbar - \hat{b}\zq.
    \]
    Then by Lemma~\ref{lemma:cascade_top}, we have the bound
    \begin{align*}
        \zq(b+6)
            &> \zq(b+5.5)e^{-(\hat{f}+\hat{b})\frac{1}{2}}+\hat{p}\left(\frac{\hat{f}}{\hat{f}+\hat{b}}\right)\left(1-e^{-(\hat{f}+\hat{b})\frac{1}{2}}\right)\\
            &> \hat{p}\frac{\hat{f}}{\hat{f}+\hat{b}}-\hat{p}e^{-\frac{\hat{f}}{2}}
            > \hat{p}\left(1-\frac{\hat{b}}{\hat{f}}\right)-\hat{p}e^{-\frac{\hat{f}}{2}}
            = \hat{p} - \hat{p}\left(\frac{\hat{b}}{\hat{f}}+e^{-\frac{\hat{f}}{2}}\right).
    \end{align*}
    Using the definition of \( k_1 \) from equation~\eqref{eq:k1_specification}, it is easy to show that \( k_1>\delta_k+\frac{2}{(1-\gamma)(1-\eta)}\log\left(\frac{4(1+\delta_0)}{\eta}\right) \).
    Therefore we know that
    \begin{align*}
        \zq(b+6)
            &> \hat{p} - \hat{p}\frac{\hat{b}}{\hat{f}}+\frac{\eta}{4}.
    \end{align*}
    It is routine but easy to show that \( \hat{p}\frac{\hat{b}}{\hat{f}}<\frac{\eta}{4} \), therefore we have the bound
    \[
        \zq(b+6) > \hat{p} - \frac{\eta}{2}.
    \]
    Finally, we know that \( \Zq \) is bounded during the interval \( [b+6,b+12] \), \( \Zq \) by
    \begin{align*}
        \D{\zq}
            &> - \hat{b}\hat{p}
            = - (k_1+\delta_k)\gamma\pzq
            > - (k_1+\delta_k)\gamma(1+\delta_0).
    \end{align*}
    It is also routine but easy to verify that \( (k_1+\delta_k)\gamma(1+\delta_0)<\frac{\eta}{12} \), and therefore \( \D{\zq} > -\frac{\eta}{12} \) during this interval.
    This means at most \( \frac{\eta}{2} \) of \( \Zq \) can be destroyed by time \( b+12 \), and thus, for all \( t\in[b+8,b+12] \), \( \zq(t)>\pzq-\eta>\pzq-\frac{1}{20} \).

    It remains to be shown that if \( q\not\in\deltaof{wa} \), then for all \( t\in[b+8,b+12] \) \( \zq(t)<\frac{1}{20} \).
    Let \( q\not\in\deltaof{wa} \).
    Then for all \( (s,a,q)\in\deltaof{wa} \), \( s\not\in\deltaof{w} \).
    Therefore, we have the following bound for \( \Zq \) in the interval \( [b+4,b+12] \).
    \begin{align*}
       \D{\zq}
            &= -\khat_1\xr\zq + \sum_{(s,a,q)\in\Delta}\khat_1\xa\ys\zqbar
            < \sum_{(s,a,q)\in\Delta}\khat_1\xa\ys\zqbar\\
            &< |Q|(k_1+\delta_k)(1+\delta_0)\eta(1+\delta_0)
            = |Q|(k_1+\delta_k)(1+\delta_0)^2\eta.
    \end{align*}
    Therefore
    \begin{align*}
        \zq(b+12)
            &< \zq(b+4) + 8|Q|(k_1+\delta_k)(1+\delta_0)^2\eta\\
            &< \eta + 8|Q|(k_1+\delta_k)(1+\delta_0)^2\eta
            < \frac{1}{20}.\qedhere
    \end{align*}
\end{proof}

\subsection{Proof of Copy Back Lemma}
\begin{proof}[Proof of Lemma~\ref{lemma:copyback}]
    In this proof, we assume the result of Lemma~\ref{lemma:transition}, \ie, \( Z_Q \) \( \frac{1}{20} \)-encodes \( \deltaof{wa} \) during the interval \( [b+8,b+12] \).
    We now focus on the behavior of \( \Yq \) during the interval \( [b+8, b+12] \).
    We begin by examining the ODE for \( \Yq \) which is
    \[
        \D{\yq} = A + B,
    \]
    where \( A=\khat_2\xc\zq\yqbar - \khat_2\xc\zqbar\yq \) and \( B=\khat_2\yq^2\yqbar - \khat_2\yq\yqbar^2 \).
    We begin by bounding the the signal restoration part of the ODE with
    \begin{align*}
        B
            &> \yq\yqbar\big((k_2-\delta_k)\yq - (k_2+\delta_k)\yqbar\big)\\
            &= -k_2\yq\yqbar\left(2\yqbar - \pyq\left(1-\frac{\delta_k}{k_2}\right)\right).
    \end{align*}
    Since \( \yq+\yqbar=\pyq \), it is not difficult to show that minimizing \( B \) under these constraints yields the inequality
    \[
        B > -\frac{k_2}{6}\pyq^3\left(1+\frac{\delta_k}{k_2}\right)^3.
    \]
    Now let \( q\in\deltaof{wa} \).
    Then by Lemma~\ref{lemma:transition}, \( \zq(t)>\pzq-\frac{1}{20} \) for all \( t\in[b+8,b+12] \).
    During the interval \( [b+9,b+10] \), the input species \( X_c \) is at a peak which means that \( X^\ast_c \) is above \( 1-\gamma \) during the interval \( [b+9.5,10] \).
    Therefore the derivative for \( \Yq \) during this interval is bounded by
    \begin{align*}
        \D{\yq}
            &> \khat_2\xc\zq\yqbar - \khat_2\xc\zqbar\yq -\frac{k_2}{6}\pyq^3\left(1+\frac{\delta_k}{k_2}\right)^3\\
            &= \hat{a}(\hat{p}-\yq) - \hat{b}\yq - \hat{c}
    \end{align*}
    where
    \begin{align*}
        \hat{a} &= k_2\left(1-\frac{\delta_k}{k_2}\right)(1-\gamma)\left(1-\delta_0-\frac{1}{20}\right)\\
        \hat{b} &= k_2\left(1+\frac{\delta_k}{k_2}\right)(1+\delta_0)\frac{1}{20}\\
        \hat{c} &= \frac{k_2}{6}\pyq^3\left(1+\frac{\delta_k}{k_2}\right)^3\\
        \hat{p} &= \pyq.
    \end{align*}
    This ODE is easily solvable, and therefore we obtain the bound
    \begin{equation}\label{eq:last_bound}
        \yq(b+9.75) > \frac{\hat{p}\hat{a}-\hat{c}}{\hat{a}+\hat{b}}\left(1-e^{-(\hat{a}+\hat{b})\frac{1}{4}}\right).
    \end{equation}
    It is routine but easy to verify that
    \begin{equation}\label{eq:last_inequality}
        \frac{\hat{p}\hat{a}-\hat{c}}{\hat{a}+\hat{b}}
            > \frac{2}{3}\pyq\left(1+\frac{\delta_k}{k_2}\right).
    \end{equation}
    Using the specification of \( k_2 \) from equation~\eqref{eq:k2_specification}, it is clear that \( k_2>\delta_k+\frac{4\log4}{(1-\gamma)(1-\delta_0-\frac{1}{20})} \).
    Therefore we know that
    \begin{align*}
        1-e^{-(\hat{a}+\hat{b})\frac{1}{4}}
            &> 1-e^{-\hat{a}\frac{1}{4}}
            = \frac{3}{4}.
    \end{align*}
    Plugging this inequality along with equation~\eqref{eq:last_inequality} into equation~\eqref{eq:last_bound}, we obtain
    \begin{align*}
        \yq(b+9.75)
            &> \frac{2}{3}\pyq\left(1+\frac{\delta_k}{k_2}\right)\left(\frac{3}{4}\right)
            = \frac{\pyq}{2}\left(1+\frac{\delta_k}{k_2}\right).
    \end{align*}
    During the interval \( [b+9.75,10] \), the derivative of \( \Yq \) is still bounded by
    \begin{align*}
        \D{\yq}
            &> \khat_2\xc\zq\yqbar - \khat_2\xc\zqbar\yq -k_2\yq\yqbar\left(2\yqbar - \pyq\left(1-\frac{\delta_k}{k_2}\right)\right)\\
            &= \khat_2\xc\zq\yqbar - \khat_2\xc\zqbar\yq +k_2\yq\yqbar\left(2\yq - \pyq\left(1+\frac{\delta_k}{k_2}\right)\right).
    \end{align*}
    Since the concentration of \( \Yq \) is greater than \( \frac{\pyq}{2}\left(1+\frac{\delta_k}{k_2}\right) \) at time \( t=9.75 \), we know that the ODE of \( \Yq \) during the interval \( [b+9.75,b+10] \) is bounded by
    \begin{align*}
        \D{\yq}
            &> \khat_2\xc\zq\yqbar - \khat_2\xc\zqbar\yq
            > \hat{a}(\hat{p}-\yq) - \hat{b}\yq.
    \end{align*}
    By Lemma~\ref{lemma:cascade_top} we obtain the bound
    \begin{align*}
        \yq(b+10)
            &> \hat{p}\left(\frac{\hat{a}}{\hat{a}+\hat{b}}\right)\left(1-e^{-(\hat{a}+\hat{b})\frac{1}{4}}\right)+\yq(b+9.75)\cdot e^{-(\hat{a}+\hat{b})\frac{1}{4}}\\
            &> \hat{p}\left(\frac{\hat{a}}{\hat{a}+\hat{b}}\right)\left(1-e^{-\hat{a}\frac{1}{4}}\right)
            > \hat{p}\left(1-\frac{\hat{b}}{\hat{a}}\right)\left(1-e^{-\hat{a}\frac{1}{4}}\right)\\
            &> \hat{p}-\hat{p}\frac{\hat{b}}{\hat{a}}-\hat{p}e^{-\hat{a}\frac{1}{4}}.
    \end{align*}
    It is routine but easy to verify that \( \hat{p}\frac{\hat{b}}{\hat{a}} < \frac{1}{40} \) and \( \hat{p}e^{-\hat{a}\frac{1}{4}} < \frac{1}{40} \), so we have the bound \( \yq(b+10)>\pyq-\frac{1}{20} \).

    It remains to be shown that if \( q\not\in\deltaof{wa} \) that \( \yq(b+10)<\frac{1}{20} \).
    This holds by symmetry of \( \Yq \) and \( \Yqbar \).
\end{proof}

\end{document}